\documentclass[%
 reprint,
 amsmath,amssymb,
 aps,
]{revtex4-2}

\usepackage{graphicx,color}
\usepackage{subcaption}
\usepackage{dcolumn}
\usepackage{bm}
\usepackage{multirow}

\usepackage{listings}



\newcommand{\revise}[1]{{\color{black}{#1}}}

\begin{document}

\preprint{APS/123-QED}

\title{Environment-adaptive machine learning potentials}



\author{Ngoc Cuong Nguyen}
\affiliation{%
Department of Aeronautics and Astronautics, Massachusetts Institute of Technology \\ 77 Massachusetts Avenue, Cambridge, MA 02139
}%

\author{Dionysios Sema}
\affiliation{%
Department of Mechanical Engineering, Massachusetts Institute of Technology \\ 77 Massachusetts Avenue, Cambridge, MA 02139
}


\date{\today}

\begin{abstract}


\revise{The development of interatomic potentials that can accurately capture a wide range of physical phenomena and diverse environments is of significant interest, but it presents a formidable challenge. This challenge arises from the numerous structural forms, multiple phases, complex intramolecular and intermolecular interactions, and varying external conditions. In this paper, we present a method to construct environment-adaptive interatomic potentials by adapting to the local atomic environment of each atom within a system. The collection of atomic environments of interest is partitioned into several clusters of atomic environments. Each cluster represents a distinctive local environment and is used to define a corresponding local potential.  We introduce a many-body many-potential expansion to smoothly blend these local potentials to ensure global continuity of the potential energy surface. This is achieved by computing the probability functions that determine the likelihood of an atom belonging to each cluster. We apply the environment-adaptive machine learning potentials to predict observable properties for Ta element and InP compound, and compare them with density functional theory calculations.}

\end{abstract}

\maketitle


\section{\label{sec:level1} Introduction}

Molecular dynamics (MD) simulations require an accurate calculation of energies and forces to analyze the physical movements of atoms. While electronic structure calculations provide accurate energies and forces, they are restricted to analyzing small length scales and short time scales due to their high computational complexity. Interatomic potentials represent the potential energy surface (PES) of an atomic system as a function of atomic positions and thus leave out the detailed electronic structures. They can enable MD simulations of large systems with millions or even billions of atoms over microseconds. 


Over the years,  empirical interatomic potentials (EIPs) such as  the Finnis-Sinclair potential \cite{Finnis1984}, embedded atom method (EAM) \cite{Daw1984}, modified EAM (MEAM) \cite{Baskes1992}, Stillinger-Weber (SW) \cite{Stillinger1985}, Tersoff \cite{Tersoff1988}, Brenner \cite{Brenner2002},  EDIP \cite{Bazant1997}, COMB \cite{Shan2010}, ReaxFF \cite{VanDuin2001} have been developed to treat a wide variety of atomic systems with different degrees of complexity. EAM potential has its root from the Finnis-Sinclair potential \cite{Finnis1984} in which the embedding function is a square root function. The MEAM potential \cite{Baskes1992}  was developed as a generalization of the EAM potential by including angular-dependent interactions in the electron density term.  The SW potential takes the form of a three-body potential in which the total energy is expressed as a linear combination of two- and three-body terms. The Tersoff potential is fundamentally different from the SW potential in that the strength of individual pair interactions is affected by the presence of surrounding atoms. The Brenner potential is based directly on the Tersoff potential but has additional terms and parameters which allow it to better describe various chemical environments. EDIP is designed to more accurately represent interatomic interactions by considering the effects of the local atomic environment on these interactions. 
\revise{Because EAM, MEAM, Tersoff, Brenner, EDIP, ReaxFF and COMB potentials dynamically adjust the strength of the bond based on the local environment of each atom, they can describe several different bonding states and  complex behaviors of atoms in various states, including defects, phase transitions, surfaces, and interfaces within materials.  One of the key features of ReaxFF and COMB is their ability to handle charge equilibration in a manner that includes long-range electrostatic interactions and reflects changes in the electronic environment of atoms during chemical reactions.}

The past decade has seen a tremendous interest in machine learning interatomic potentials (MLIPs) due to their promising quantum accuracy at significantly lower computational complexity than electronic structure calculations. The descriptors play a central role in the construction of accurate and efficient MLIPs. In recent years, a wide variety of descriptors has been developed to represent atomic structures. There are two main approaches to mapping a configuration of atoms onto descriptors \cite{Musil2021}: atom density approach and internal coordinate approach. Examples of internal coordinate descriptors include permutation-invariant polynomials (PIPs) \cite{Braams2009,Nguyen2018a,VanDerOord2020}, atom-centered symmetry functions (ACSFs) \cite{Behler2007,Behler2011,Behler2014}, and proper orthogonal descriptors (PODs) \cite{Nguyen2023,Rohskopf2023,nguyen2024b}. These internal coordinate descriptors are intrinsically invariant with respect to translation and rotation because they are functions of angles and distances. They are made to be permutationally invariant by summing symmetry functions over all possible atomic pairs and triplets within local atomic environments. However, achieving permutation invariance by such a way leads to the exponential scaling in terms of the number of neighbors. The computational cost can be kept under control by restricting the range of interactions, the number of descriptors, and the body orders. 

The atom density approach describes a local atomic environment around a central atom as an atom density function which is obtained by summing over localized functions centered on the relative positions of all atoms in the local environment. Such a density is naturally invariant to translation and permutation. The atomic neighborhood density is then expanded as a linear combination of appropriate basis functions, where the expansion coefficients are given by the inner products of the neighborhood density with the basis functions. Rotationally invariant descriptors are computed as appropriate sums of products of the density coefficients. In the atom density approach, the choices of the basis set (e.g., radial basis functions, spherical harmonics, angular monomials, hyperspherical harmonics) lead to different sets of descriptors. The power spectrum and bispectrum descriptors \cite{Bartok2013} are constructed from spherical harmonics, while the spectral neighbor analysis potential (SNAP) descriptors \cite{Thompson2015} are based on hyperspherical harmonics. The moment tensor potential (MTP)  \cite{Shapeev2016} projects the atomic density onto a tensor product of angular vectors to construct the moment tensors whose contraction results in invariant descriptors. The atomic cluster expansion (ACE) \cite{Drautz2019,Drautz2020} extends the power and bispectrum construction to obtain a complete set of invariant descriptors with arbitrary number of body orders. The E3 equivariant graph neural network potentials \cite{Batzner2022} use spherical harmonics. The atom density representation of POD descriptors employs angular monomials and radial basis functions constructed from the proper orthogonal decomposition \cite{Nguyen2023b}. 

The main advantage of atom density descriptors is that their computational complexity scales linearly with the number of neighbors irrespective of the body orders. The  computational complexity of internal coordinate descriptors  scales exponentially with the body order in terms of the number of neighbors. However, the cost of internal coordinate descriptors scales linearly with the number of basis functions, whereas that of atom density descriptors scales exponentially with the body order in terms of the number of basis functions. In general, atom density descriptors  are more efficient than  internal coordinate descriptors when there are many neighbors and the body order is higher than 3. Despite the rather fundamental difference in their construction, some internal coordinate descriptors and atom density descriptors can be shown to span the same descriptor space. This is the case for the POD descriptors in which the atom density representation is shown to be equivalent to the internal coordinate representation \cite{Nguyen2023b}. The POD formalism allows other internal coordinate descriptors like PIPs and ACSFs, as well as empirical potentials like EAM, MEAM, and SW, to be implemented using the atom density approach. 

\begin{figure*}
\includegraphics[width=0.8\textwidth]{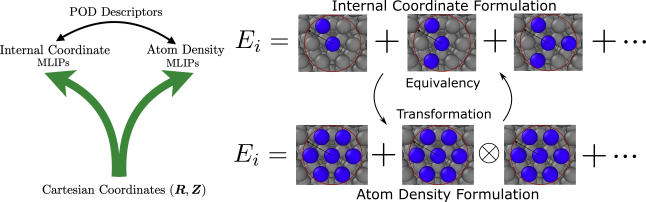}
\caption{\label{figIDEF1} Proper Orthogonal Descriptors can be formulated with either the internal coordinate or atom density representations with a transformation that shows their equivalency.}
\end{figure*}
 
Despite considerable progress that has been made in recent years, there remain open problems to be addressed with regard to the accuracy, efficiency, and transferability of interatomic potentials. The development of interatomic potentials that can effectively capture a wide range of atomic environments is a complex challenge due to several reasons. Materials can exist in numerous structural forms (e.g., crystalline, amorphous, defects, interfaces) and phases (solid, liquid, gas, plasma). Atoms interact through various forces such as electrostatic, van der Waals, ionic bonding, covalent bonding, and metallic bonding, which manifest differently depending on the chemical elements and their electronic structures.  Furthermore, the effective interaction among atoms can change with external conditions like temperature, pressure, and chemical environment. Consequently, creating an interatomic potential that performs well across diverse conditions is difficult because optimizing the potential for one set of conditions  can lead to poorer performance in others. Each of these factors contributes to the complexity of developing interatomic potentials that are effective and efficient to capture a diverse range of local atomic environments. 

In this paper, we introduce a  method for the systematic construction of accurate and transferable interatomic potentials by adapting to the local atomic environment of each atom within a system. Local atomic environment of an atom comprises the positions and chemical species of the atom and its neighbors within a cutoff radius. These atom positions and chemical species can be mapped onto a vector of $M$ invariant descriptors by using either the internal coordinate approach or the atom density approach. For a dataset of $N$ atoms, we obtain a descriptor matrix of size $N$ by $M$. Each row of the descriptor matrix encapsulates the local atomic environment of the corresponding atom. Since $M$ is typically large, a dimensionality reduction technique is used to compress the descriptor matrix into a lower-dimensional matrix of size $N$ by $J$, where $J$ is considerably less than $M$. A clustering method is then employed to partition the compressed data into $K$ separate clusters. In other words, the original dataset of $N$ atoms is divided into $K$ subsets and the atoms in any subset have similar atomic environments. The clustering scheme allows us to divide the diverse dataset into smaller subsets, each characterized by  data points sharing the common attributes. This approach captures the diversity inherent in the dataset by identifying distinct atomic environments within the dataset. By training MLIPs on these subsets separately, we can obtain MLIPs that are tailored to specific atomic environments. Each MLIP may accurately predict configurations in the subset on which it is trained. However, it may not be accurate for predicting configurations in the other subsets. 

The above approach raises the question: \revise{How do we combine these separately localized MLIPs to construct a global potential energy surface?}  To this end, we propose a many-body many-potential (MBMP) expansion designed to seamlessly blend the individual MLIPs and ensure that the potential energy surface remains continuous across cluster boundaries. This continuity is achieved by calculating probability functions that assess the likelihood of an atom belonging to specific clusters identified within the dataset. These probability functions are critical in guiding how contributions from different MLIPs are weighted and combined, providing a systematic way to maintain the integrity and accuracy of the model across different atomic environments. This integration is crucial for achieving a comprehensive model that can accurately capture diverse environments in the original dataset. \revise{This model can also capture atomic environments that are a mixture of several distinct environments when the probability functions are close to each other, thereby potentially making the model more transferable than the individual MLIPs.}

Although the formulation of the environment-adaptive machine learning (EAML) potentials is descriptor agnostic and can be developed for any set of descriptors, in this work we employ the proper orthogonal descriptors \cite{Nguyen2023,Nguyen2023b}. To this end, we extend the proper orthogonal descriptors to deal with multi-element systems. This  enables us to construct EAML potentials that are finely tuned to the complexities of various material compositions under diverse conditions. We apply the EAML potentials to predict observable properties for Ta element and InP compound, and compare them with density functional theory calculations.

The paper is organized as follows. In Section \ref{fastpod}, we extend proper orthogonal descriptors to multi-element systems. In Section \ref{EAMLmethods}, we describe our approach for constructing EAML potentials. In Section \ref{results}, we present results to demonstrate the EAML potentials for Tantalum and Indium Phosphide. Finally, we provide some concluding remarks in Section \ref{conclusions}.

\section{Multi-Element proper orthogonal descriptors}
\label{fastpod}

\revise{This section outlines a systematic approach for constructing internal coordinate and atom density descriptors to represent the local atomic environments of multi-element systems. Building on our previous work \cite{Nguyen2023,Rohskopf2023,nguyen2024b}, we develop invariant descriptors for multi-element systems by leveraging orthogonal proper decomposition to generate radial basis functions and employing the trinomial expansion of angular monomials to achieve rotational symmetry. The resulting descriptors combine elements of both internal coordinates and atom density fields, as illustrated in Figure \ref{figIDEF1}.}

\subsection{Many-body potential energy surface}

We consider a multi-element system of $N_{a}$ atoms with $N_{e}$ unique elements. We denote by $\bm r_i$ and $Z_i$ position vector and type of an atom $i$ in the system, respectively. Thus we have $Z_i \in \{1, \ldots, N_{e} \}$, $\bm R = (\bm r_1, \bm r_2, \ldots, \bm r_{N_{a}}) \in \mathbb{R}^{3N_{a}}$, and $\bm Z = (Z_1, Z_2, \ldots, Z_{N_{a}}) \in \mathbb{N}^{N_{a}}$. The potential energy surface (PES) of the system of $N_{a}$ atoms can be expressed as a many-body expansion of the form
\begin{equation}
\label{eq1}
\begin{split}
E_{\rm T}(\bm R, \bm Z)  =  & \sum_{i} V^{(1)}(\bm r_i, Z_i) +  \sum_{i,j} V^{(2)}(\bm r_i, \bm r_j, Z_i, Z_j)  \\
& + \sum_{i,j,k} V^{(3)}(\bm r_i, \bm r_j, \bm r_k,  Z_i, Z_j, Z_k)  \\ 
& + \sum_{i,j,k,l} V^{(4)}(\bm r_i, \bm r_j, \bm r_k, \bm r_l,  Z_i, Z_j, Z_k, Z_l) + \ldots 
\end{split}
\end{equation}
The superscript on each potential denotes its body order. Each potential must also depend on a set of parameters used to parametrize it for a specific application. To simplify the notation, we have chosen not to explicitly denote these parameters in the potentials. A separation of the PES into atomic contributions
yields
\begin{equation}
E_{\rm T}(\bm R, \bm Z) = \sum_{i=1}^{N_{a}} E_i(\bm R, \bm Z)
\end{equation}
where $E_i$ is obtained from (\ref{eq1}) by removing the sum over index $i$. To make the PES invariant with respect to translation and rotation, the potentials should depend only on internal coordinates as follows 
\begin{equation}
\label{eq2}
\begin{split}
E_i  =  &  V^{(1)}(Z_i)+  \sum_{j} V^{(2)}( r_{ij}, Z_i, Z_j) \ +  \\
& \sum_{j,k} V^{(3)}(r_{ij}, r_{ik}, w_{ijk}, Z_i, Z_j, Z_k) \ + \\ 
&  \sum_{j,k,l} V^{(4)}(r_{ij}, r_{ik}, r_{il}, w_{ijk}, w_{ijl}, w_{ikl}, Z_i, Z_j, Z_k, Z_l) + \ldots 
\end{split}
\end{equation}
where $\bm r_{ij} = \bm r_j - \bm r_i$, $r_{ij} = |\bm r_{ij}|$, $w_{ijk} = \cos \theta_{ijk} = \hat{\bm r}_{ij} \cdot \hat{\bm r}_{ik}$, $\hat{\bm r} = \bm r/|\bm r|$. The internal coordinates include both distances $r_{ij}, r_{ik}, r_{il}$ and angles $w_{ijk}, w_{ijl}, w_{ikl}$. The number of internal coordinates for $V^{(q)}$ is equal to $(q-1)q/2$. Typically, the one-body terms $V^{(1)}(Z_i)$ are set to the isolated energies of atom $i$.

\subsection{Two-body proper orthogonal descriptors}
\label{twobodypod}

We briefly describe the construction of two-body PODs and refer to \cite{Nguyen2023,Nguyen2023b} for further details. We assume that the direct interaction between two atoms vanishes smoothly when their distance is greater than the cutoff distance $r_{\rm cut}$. Furthermore, we assume that two atoms can not get closer than the inner cutoff distance $r_{\rm in}$. Letting $r \in (r_{\rm in}, r_{\rm cut})$, we introduce the following parametrized radial functions
\begin{equation}
\label{eq3}
\phi(r, r_{\rm in}, r_{\rm cut}, \alpha, \beta)  =  \frac{\sin (\alpha \pi x) }{\alpha(r - r_{\rm in})}, \qquad  \varphi(r, \gamma)  = \frac{1}{r^\gamma} ,    
\end{equation}
where the scaled distance function $x$ is given by
\begin{equation}
x(r, r_{\rm in}, r_{\rm cut}, \beta) = \frac{e^{-\beta(r - r_{\rm in})/(r_{\rm cut} - r_{\rm in})} - 1}{e^{-\beta} - 1} .
\end{equation}
The function $\phi$ in (\ref{eq3}) is related to the zeroth spherical Bessel function, while the function $\varphi$ is inspired by the n-m Lennard-Jones potential. Although the parameter $\gamma$ can be real number, we choose a set of consecutive positive integers  $\{1,2,\ldots,  P_\gamma\}$ to compute instances of the parametrized function $\varphi$ by making use of the relation $\varphi(r, \gamma+1) = \varphi(r, \gamma)/r$. Similarly, we choose a set of consecutive integers $\{1, 2, \ldots, P_{\alpha}\}$ for $\alpha$ to generate instances of the parametrized function $\phi$ by making use of the formula $\sin((\alpha+1) \pi x) = \sin(\pi x) U_{\alpha}(\cos (\pi x))$, where $U_{\alpha}$ are Chebyshev polynomials of the second kind. We take $P_{\beta}$ values for the parameter $\beta$ such that $\beta_k = (k-1) \beta_{\max}/(P_{\beta}-1)$ for $k = 1, 2, \ldots, P_{\beta}$, where $\beta_{\max} = 4.0$.

We introduce the following function as a convex combination of the two functions in (\ref{eq3})
\begin{equation}
\psi(r, \bm \mu)  = \kappa \phi(r, r_{\rm in}, r_{\rm cut}, \alpha, \beta) + (1- \kappa)  \varphi(r, \gamma) ,
\end{equation}
where $\mu_1 = r_{\rm in}, \mu_2 = r_{\rm cut}, \mu_3 = \alpha, \mu_4 = \beta, \mu_5 = \gamma$, and $\mu_6 = \kappa$. The two-body parametrized potential is defined as follows
\begin{equation}
\label{eq6}
V^{(2)}(r_{ij}, \bm \mu)  = f_{\rm c}(r_{ij}, \bm \mu) \psi(r_{ij}, \bm \mu)
\end{equation}
where the cut-off function $f_{\rm c}(r_{ij}, \bm \mu)$ is 
\begin{equation}
\label{eq8}
 f_{\rm c}(r, \bm \mu)  =  \exp \left(1 -\frac{1}{\sqrt{\left(1 - \frac{(r-r_{\rm in})^3}{(r_{\rm cut} - r_{\rm in})^3} \right)^2 + \epsilon}} \right)
\end{equation}
with $\epsilon = 10^{-6}$. This cut-off function ensures the smooth vanishing of the two-body potential and its derivative for $r_{ij} \ge r_{\rm cut}$.

Given $S = P_\gamma + P_{\alpha} P _\beta$ parameter tuples $\bm \mu_s, 1 \le s \le S$, we introduce the following set of snapshots 
\begin{equation}
\label{eq11}
\Phi_s(r_{ij}) =  V^{(2)}(r_{ij}, \bm \mu_s),  \quad s = 1, \ldots, S .
\end{equation}
We next employ the proper orthogonal decomposition \cite{Nguyen2023} to generate an orthogonal basis set which is known to be optimal for representation of the snapshot family $\{\Phi_s\}_{s=1}^S$.  In particular, the orthogonal radial basis functions are computed as follows
\begin{equation}
\label{eq12}
R_n(r_{ij}) = \sum_{s = 1}^S Q_{s n} \,  \Phi_s(r_{ij}), \qquad n = 1, \ldots, N_r , 
\end{equation}
where the number of radial basis functions $N_r$ is typically in the range between 5 and 10. Note that $Q_{s n}, 1 \le s \le S, 1 \le n \le N_r,$ are a  matrix whose columns are eigenvectors of the following eigenvalue problem 
\begin{equation}
\bm C \bm a = \lambda \bm a, 
\end{equation} 
where the covariance matrix $\bm C$ is given by
\begin{equation}
C_{s p} = \int_{r_{\rm in}}^{r_{\rm cut}} \Phi_s(r) \Phi_p(r) d r, \qquad 1 \le s, p \le  S .  
\end{equation} 
The covariance matrix is computed by using the trapezoidal rule on a grid of 2000 subintervals on the interval $[r_{\rm in}, r_{\rm cut}]$. The eigenvector matrix $Q_{sn}$ is pre-computed and stored.

Finally, the two-body proper orthogonal descriptors at each atom $i$ are computed by summing the orthogonal basis functions over the neighbors of atom $i$ and numerating on the atom types as follows
\begin{equation}
\label{eq12b}
\mathcal{D}^{(2)}_{ip q n  }  = \left\{
\begin{array}{ll}
\displaystyle \sum_{\{j=1 | Z_j = q\}}^{N_i} R_n(r_{ij}), & \quad \mbox{if } Z_i = p \\
0, & \quad \mbox{if } Z_i \neq p
\end{array} 
\right.   
\end{equation}
for $1 \le i \le N_{a}, 1 \le n \le N_{r}, 1 \le q, p \le N_{e}$. The number of two-body descriptors per atom is thus $N_{r} N_{e}^2$. 

\revise{For the purpose of complexity analysis, we assume that each atom has the same number of neighbors $N_i$. The total number of neighbors is thus $N_a N_i$ for all atoms. The cost of evaluating the radial basis functions in (\ref{eq12}) is $O(N_a N_i N_r S)$, while the cost of evaluating the two-body descriptors in (\ref{eq12b}) is $O(N_a N_i N_r)$. The total cost is thus independent of the number of elements $N_{e}$.} 

\subsection{Three-body proper orthogonal descriptors}

For any given integer $\ell \in [0,P_a]$, where $P_a$ is the highest angular degree, we introduce a basis set of angular monomials 
\begin{equation}
\label{eq23b}
A_{\ell m} (\hat{\bm r}_{ij} ) = (\hat{x}_{ij})^{l_x} \ (\hat{y}_{ij})^{l_y} \ (\hat{z}_{ij})^{l_z}, 
\end{equation}
where the exponents $l_x, l_y, l_z$ are nonnegative integers such that ${l_x} + l_y + l_z = \ell$.  Note that the index $m$ satisfies $0 \le m \le (\ell+1)(\ell+2)/2 - 1$ for any given $\ell$, and that the total number of angular monomials is $(P_a+1)(P_a+2)(P_a+3)/6$. Table \ref{table3} shows the basis set of angular monomials for $P_a = 4$. By applying the trinomial expansion to the power of the angle component $w_{ijk}$, we obtain
\begin{equation}
\begin{split}
(w_{ijk})^{\ell} & = \left( \hat{x}_{ij} \hat{x}_{ik} + \hat{y}_{ij} \hat{y}_{ik} + \hat{z}_{ij} \hat{z}_{ik} \right)^{\ell} \\
&= \sum_{m=0}^{L(\ell)} C_{\ell m} A_{\ell m}(\hat{\bm r}_{ij}) A_{\ell m}(\hat{\bm r}_{ik})
\end{split}
\end{equation}
where $L(\ell) = (\ell+1)(\ell+2)/2-1$ and $C_{\ell m} = \frac{\ell!}{l_x! l_y! l_z!}$ correspond to the multinomial coefficients of the trinomial expansion. Table \ref{table4} shows the multinomial coefficients for $\ell=0,1,2,3,4$.

\begin{table}[htbp]
\caption{\label{table3}
The angular monomials for $P_a = 4$.
}
\begin{ruledtabular}
\begin{tabular}{ll}
$\ell$ &
$A_{\ell m}(\hat{\bm r}_{ij} )$ \\
\colrule
0 & $1$ \\
1 & $\hat{x}_{ij}$, \ $\hat{y}_{ij}$, \ $\hat{z}_{ij}$ \\[0.5ex]
2 & $\hat{x}^2_{ij}$, \ $\hat{y}^2_{ij}$, \ $\hat{z}^2_{ij}$, \ $\hat{x}_{ij} \hat{y}_{ij}$, \ $\hat{x}_{ij} \hat{z}_{ij}$, \ $\hat{y}_{ij} \hat{z}_{ij}$ \\[0.5ex]
3 & $\hat{x}^3_{ij}$, \ $\hat{y}^3_{ij}$, \ $\hat{z}^3_{ij}$, \ $\hat{x}^2_{ij} \hat{y}_{ij}$, \ $\hat{x}^2_{ij} \hat{z}_{ij}$, \ $\hat{y}^2_{ij} \hat{x}_{ij}$, \ $\hat{y}^2_{ij} \hat{z}_{ij}$,  \\[0.5ex] & $\hat{z}^2_{ij} \hat{x}_{ij}$, $\hat{z}^2_{ij} \hat{y}_{ij}$, $\hat{x}_{ij} \hat{y}_{ij} \hat{z}_{ij}$ \\[0.5ex]  
4 & $\hat{x}^4_{ij}$, \ $\hat{y}^4_{ij}$, \ $\hat{z}^4_{ij}$, \ $\hat{x}^3_{ij} \hat{y}_{ij}$, \ $\hat{x}^3_{ij} \hat{z}_{ij}$, \ $\hat{y}^3_{ij} \hat{x}_{ij}$, \ $\hat{y}^3_{ij} \hat{z}_{ij}$, $\hat{z}^3_{ij} \hat{x}_{ij}$  \\[0.5ex] 
& $\hat{z}^3_{ij} \hat{y}_{ij}$, $\hat{x}^2_{ij} \hat{y}^2_{ij}$, \ $\hat{x}^2_{ij} \hat{z}^2_{ij}$, \ $\hat{y}^2_{ij} \hat{z}^2_{ij}$, \ $\hat{x}^2_{ij} \hat{y}_{ij} \hat{z}_{ij}$ \ $\hat{x}_{ij} \hat{y}^2_{ij} \hat{z}_{ij}$ \\[0.5ex]  
& $\hat{x}_{ij} \hat{y}_{ij} \hat{z}^2_{ij}$ \\[0.5ex]  
\end{tabular}
\end{ruledtabular}
\end{table}

\begin{table}[htbp]
\caption{\label{table4}
The multinomial coefficients for $P_a = 4$.
}
\begin{ruledtabular}
\begin{tabular}{ll}
$\ell$ &
$C_{ \ell m}$ \\
\colrule
0 & $1$ \\
1 & $1$, \ $1$, \ $1$ \\[0.5ex]
2 & $1$, \ $1$, \ $1$, \ $2$, \ $2$, \ $2$   \\[0.5ex]
3 & $1$, \ $1$, \ $1$, \ $3$, \ $3$, \ $3$, \ $3$, \ $3$, \ $3$, \ $6$    \\[0.5ex]
4 & $1$, \ $1$, \ $1$, \ $4$, \ $4$, \ $4$, \ $4$, \ $4$, \ $4$, \ $6$, \ $6$, \ $6$, \ $12$, \ $12$, \ $12$    \\[0.5ex]
\end{tabular}
\end{ruledtabular}
\end{table}

Next, we define the atom basis functions at atom $i$ as the sum over all neighbors of atom $i$ of the products of radial basis functions and angular monomials  
\begin{equation}
\label{24b}
B_{i q n \ell m}  = \sum_{j=1 | Z_j = q}^{N_i} R_n(r_{ij})  A_{\ell m}(\hat{\bm r}_{ij}), \quad 1 \le q \le  N_{e} .
\end{equation}
The cost of evaluating the atom basis functions is $O(N_a N_i N_r (P_a+1) (P_a+2) (P_a+3)/6 )$, which is independent of the number of elements. These atom basis functions are used to define the atom density descriptors as follows 
\begin{equation}
\label{eq26a}
\mathcal{D}^{(3)}_{ip q q' n \ell }  = \left\{
\begin{array}{ll}
\displaystyle \sum_{m=0}^{L(\ell)} C_{\ell m} B_{i q n \ell m}  B_{i q' n \ell m} , & \quad \mbox{if } Z_i = p \\
0, & \quad \mbox{if } Z_i \neq p
\end{array} 
\right.   
\end{equation}
for $1 \le i \le N_{a}, 1 \le n \le N_{r}, 0 \le \ell \le P_a, 1 \le q, p \le N_{e}, 1 \le q' \le q$. The number of three-body descriptors per atom is thus $N_{r} (P_{a}+1) N_{e}^2 (N_{e} + 1)/2$. The cost of evaluating (\ref{eq26a}) is $O(N_a N_{r} N_e (N_e+1) (P_a+1)(P_a+2)(P_a+3)/12)$, which is usually less than that of evaluating the atom basis functions. Therefore, the total cost of computing the atom density descriptors is $O(N_a N_i N_r (P_a+1) (P_a+2) (P_a+3)/6 )$.

Substituting (\ref{24b}) into (\ref{eq26a}) yields the internal coordinate form of the three-body descriptors 
\begin{equation}
\label{eq26d}
\mathcal{D}^{(3)}_{ip q q' n \ell }  = 
\displaystyle \sum_{j|Z_j = q}^{N_i} \sum_{k|Z_k = q'}^{N_i} R_n(r_{ij})  R_n(r_{ik}) \left( w_{ijk} \right)^{\ell} 
\end{equation}
for $Z_i = p$. The cost of evaluating the internal coordinate descriptors is $O(N_a N^2_{i} N_{r} P_{a})$. Although the atom density descriptors (\ref{eq26a}) and the internal coordinate descriptors (\ref{eq26d}) are mathematically equivalent, their computational complexity are not the same. The complexity of the atom density descriptors is linear in the number of neighbors and cubic in the angular degree, whereas that of the internal coordinate descriptors is quadratic in the number of neighbors and linear in the angular degree. Therefore, if the number of neighbors is large and the angular degree is small, it is faster to evaluate the the atom density descriptors. On the other hand, if the number of neighbors is small and the angular degree is high, it is more efficient to compute the internal coordinate descriptors.

\subsection{Four-body proper orthogonal descriptors}

We begin by introducing the following four-body angular functions
\begin{equation}
\label{eq16a}
f_{s}(w_{ijk}, w_{ijl}, w_{ikl}) = \left( w_{ijk} \right)^{a} \left( w_{ijl} \right)^{b} \left( w_{ikl} \right)^{c},
\end{equation}
where $a,b,c$ are integers such that $a + b + c = \ell$ and $a \ge b \ge c \ge 0$. The four-body angular functions $f_s$ are listed in Table \ref{table1}. The four-body internal coordinate descriptors at each atom $i$ are defined as
\begin{equation}
\label{eq18a}
\mathcal{D}^{(4)}_{ip q q' q'' ns}  = \displaystyle \sum_{ \{j|Z_j = q \}}^{N_i} \sum_{\{k|Z_k = q'\}}^{N_i} \sum_{\{l|Z_l = q'' \}}^{N_i}  U_{ns} 
\end{equation}
for $Z_i = p$, where $U_{ns}$ are given by
\begin{equation}
\label{eq17a}
U_{ns} = R_n(r_{ij}) R_n(r_{ik}) R_n(r_{il}) f_s(w_{ijk}, w_{ijl}, w_{ikl})
\end{equation}
for $1 \le i \le N_{a}, 1 \le n \le N_{r}, 1 \le s \le K_a, 1 \le p, q \le N_{e}, 1 \le q' \le q, 1 \le q'' \le q'$. Here $K_a$ is the number of four-body angular basis functions, which depends on $P_a$. The number of four-body descriptors per atom is thus $N_{r} K_{a} N_{e}^2 (N_{e} + 1)(N_{e} + 2)/6$. The cost of evaluating the four-body internal coordinate descriptors is $O(N_a N^3_{i} N_{r} K_{a})$, which is independent of the number of elements.

\begin{table}[htbp]
\caption{\label{table1}%
Four-body angular functions $f_s$
}
\begin{ruledtabular}
\begin{tabular}{ll}
$\ell$ &
$f_s$ \\
\colrule
0 & $f_1 = 1$   \\
1 & $f_2 = w_{ijk}$\\[0.5ex]
2 & $f_3 = w^2_{ijk}, \quad f_4 = w_{ijk} w_{ijl}$ \\[0.5ex]
3 & $f_5 = w^3_{ijk}, \quad f_6 = w^2_{ijk} w_{ijl}$, \quad $f_7 = w_{ijk} w_{ijl} w_{ikl}$ \\[0.5ex]
4 & $f_8 = w^4_{ijk}, \quad f_9 = w^3_{ijk} w_{ijl}$, \quad $f_{10} = w^2_{ijk} w^2_{ijl}$, \\[0.5ex]
& $f_{11} = w^2_{ijk} w_{ijl} w_{ikl}$ \\
\end{tabular}
\end{ruledtabular}
\end{table}

We note from the trinomial expansion that
\begin{equation*}
\begin{split}
(\xi_1 & + \xi_2 + \xi_3)^a  (\eta_1 + \eta_2 + \eta_3)^b (\zeta_1 + \zeta_2 + \zeta_3)^c  = \\
& \sum_{\alpha =0}^{L(a)} \sum_{\beta=0}^{L(b)} \sum_{\gamma=0}^{L(c)} C_{a \alpha} C_{b \beta} C_{c \gamma} A_{a \alpha}(\bm \xi)  A_{b \beta}(\bm \eta) A_{c \gamma}(\bm \zeta) 
\end{split}
\end{equation*}
where $A_{a\alpha}$ are the angular monomials defined in (\ref{eq23b}). By considering $\xi_1 = \hat{x}_{ij} \hat{x}_{ik}, \xi_2 = \hat{y}_{ij} \hat{y}_{ik}, \xi_3 = \hat{z}_{ij} \hat{z}_{ik}$, $\eta_1 = \hat{x}_{ij} \hat{x}_{il}, \eta_2 = \hat{y}_{ij} \hat{y}_{il}, \eta_3 = \hat{z}_{ij} \hat{z}_{il}$, $\zeta_1 = \hat{x}_{ik} \hat{x}_{il}, \zeta_2 = \hat{y}_{ik} \hat{y}_{il}, \zeta_3 = \hat{z}_{ik} \hat{z}_{il}$, we obtain
\begin{equation*}
A_{a \alpha}(\bm \xi)  A_{b \beta}(\bm \eta) A_{c \gamma}(\bm \zeta) =  A_{a' \alpha'}(\hat{\bm r}_{ij})  A_{b' \beta'}(\hat{\bm r}_{ik})  A_{c' \gamma'}(\hat{\bm r}_{il})  
\end{equation*}
where $a' = a + b$, $b' = a + c$, $c' = b + c$, and the index $\alpha'$ depends on $\alpha$ and $\beta$, $\beta'$  on $\alpha$ and $\gamma$, $\gamma'$ on $\beta$ and $\gamma$. It thus follows that the four-body angular functions can be expressed as 
\begin{equation*}
f_{s} =  \sum_{\alpha =0}^{L(a)} \sum_{\beta=0}^{L(b)} \sum_{\gamma=0}^{L(c)} C_{abc}^{\alpha \beta \gamma} A_{a' \alpha'}(\hat{\bm r}_{ij} ) A_{b' \beta'}(\hat{\bm r}_{ik} ) A_{c' \gamma'}(\hat{\bm r}_{il}) .
\end{equation*}
where $C_{abc}^{\alpha \beta \gamma} = C_{a \alpha} C_{b \beta} C_{c \gamma}$. Hence, the internal coordinate form (\ref{eq18a}) is equivalent to the  atom density form
\begin{equation}
\label{eq30a}
\begin{split}
\mathcal{D}^{(4)}_{ipqq'q''ns} = &  \sum_{\alpha =0}^{L(a)} \sum_{\beta=0}^{L(b)} \sum_{\gamma=0}^{L(c)} C_{abc}^{\alpha \beta \gamma}  B_{iq n a' \alpha'} B_{iq' n b' \beta'} B_{iq'' n c' \gamma'} .
\end{split}
\end{equation}
The four-body atom density descriptors are expressed  in terms of the sums of the products of the atom basis functions. In general, they are more efficient to evaluate than their internal coordinate counterparts because they scale linearly with the number of neighbors. The complexity analysis of the four-body atom density descriptors is detailed in \cite{Nguyen2023b}.

It is possible to exploit the symmetry and hierarchy  of the four-body descriptors to reduce the computational cost. In particular, the four-body descriptors associated with $f_1 = 1$ are the products of three two-body descriptors.  For $b = c = 0$ (e.g., $f_2, f_3, f_5, f_8$ in Table \ref{table1}) the four-body descriptors are the products of  two-body descriptors and  three-body descriptors. Those four-body descriptors can be computed very fast without using the atom density form (\ref{eq30a}). The remaining four-body descriptors are calculated by using (\ref{eq30a}). They share many common terms, which can be exploited to further reduce the cost. For instance, since $f_6 = w_{ijk} f_4$ and $f_9 = w^2_{ijk} f_4$, the descriptors associated with $f_6$ and $f_9$  have many common terms with those associated with $f_4$.

\section{Environment-Adaptive Machine Learning Potentials}
\label{EAMLmethods}

This section describes a method for constructing EAML potentials from a given set of invariant descriptors. The method leverages the principal component analysis and $k$-means algorithm to partition the dataset into atom clusters. The method relies on a many-body many-potential expansion that combines several different potentials to define a single potential energy surface. This is done by calculating probability functions that assess the likelihood of an atom belonging to specific clusters. These probability functions determine how contributions from different potentials are weighted and combined and provide a systematic way to maintain the continuity of the potential energy surface.

\subsection{Linear regression models}

Linear regression is the most simple and efficient method for building MLIP models \cite{Thompson2015,Wood2018,Novoselov2019,Shapeev2016}. Let ${\mathcal{D}}_{im}, 1 \le m \le M,$ be a set of $M$ local descriptors at atom $i$. The atomic energy at an atom $i$ is expressed as a linear combination of the local descriptors 
\begin{equation}
\label{eq57a}
E_i(\bm R, \bm Z) = \sum_{m=1}^M c_m \mathcal{D}_{im}  (\bm R, \bm Z) 
\end{equation}
where  $c_m$ are the coefficients to be determined by fitting against QM database. The PES is given by
\begin{equation}
\label{eq58a}
E_{\rm T}(\bm R, \bm Z) = \sum_{i=1}^{N_a} E_i(\bm R, \bm Z) = \sum_{m=1}^M c_m \mathcal G_{m} (\bm R, \bm Z)    
\end{equation}
where $\mathcal G_m = \sum_{i=1}^{N_a} \mathcal{D}_{im}$ are the global descriptors. The coefficients $c_m$ are sought as solution of a least squares problem 
\begin{equation}
\label{eq63a}
\min_{\bm c} \alpha \|\bm G \bm c - \bm e \|^2 + \beta \|\bm H \bm c - \bm f \|^2 + \gamma \|\bm c\|^2 
\end{equation}
where the matrix $\bm G$ is formed from the global descriptors, while $\bm H$ is formed from the derivatives of the global descriptors for all configurations in the training database. The vector $\bm e$ is comprised of  DFT energies, while $\bm f$ is comprised of DFT forces. Note that $\alpha$ is the energy weight parameter, $\beta$ is the force weight parameter, and $\gamma$ is a regularization parameter. They are hyperparameters of the linear regression model. 


In order to accurately model atomic forces in MD simulations, the training dataset must be diverse and rich enough to cover structural forms (e.g., crystalline, amorphous, defects, interfaces), multiple phases (solid, liquid, gas, plasma), and a wide range of temperature, pressure, and chemical environment. \revise{Training a linear model on the entire training set may not produce an accurate and efficient potential it the dataset contains very diverse environments.} We partition the training dataset into $K$ separate subsets, whose DFT energies and forces are denoted by $(\bm e^k, \bm f^k), 1 \le k \le K$. On each subset, we introduce an associated PES
\begin{equation}
\label{eq57b}
E^k_{\rm T}(\bm R, \bm Z) = \sum_{i=1}^{N_a} \sum_{m=1}^M c_{m}^k \mathcal{D}_{im}  (\bm R, \bm Z) , 
\end{equation}
where the coefficient vectors $\bm c^k$ are sought as solutions of the least squares problems 
\begin{equation}
\label{eq63aw}
\min_{\bm c^k} \alpha^k \|\bm G^k \bm c^k - \bm e^k \|^2 + \beta^k \|\bm H^k \bm c^k - \bm f^k \|^2 + \gamma^k \|\bm c^k\|^2 .
\end{equation}
Here the superscript $k$ is used to indicate the quantities associated with the $k$th subset. This training strategy yields an ensemble of $K$ separate potentials. Each potential may accurately predict configurations in the subset on which it is trained. However, it may not be accurate for predicting configurations in the other subsets. 

\revise{A hypothetical issue here is to guarantee  the continuity of PES when predicting forces with an ensemble of 
potentials. It is not obvious how to combine these separate potentials, as they are trained on different datasets. A simple strategy is to select the best potential among these potentials to predict the physical properties of a given configuration at hand, if the criterion of selecting the best potential can be defined. While this strategy may work for property prediction, it does not work for MD simulations. This is because using different potentials in an MD simulation will result in discontinuity in the PES and thus forces. The remainder of this section describes a method that allows us to combine these separate potentials to construct a global, differentiable and continuous PES.}


\subsection{Dataset partition}

We describe a clustering method to partition the dataset into subsets of similar attributes. Local atomic environment of an atom $i$ comprises the positions and chemical species of the atom and its neighbors within a cutoff radius. These atom positions and chemical species can be mapped onto a vector of $M$ invariant descriptors, $\mathcal{D}_{im}, 1 \le m \le M$, by using either the internal coordinate approach or the atom density approach. For a dataset of $N$ atoms, we obtain a descriptor matrix $\bm{\mathcal{D}}$ of size $N$ by $M$. Each row of the descriptor matrix encapsulates the local atomic environment of the corresponding atom. The similarity between two local atomic environments can be measured by the dot product of the two corresponding descriptor vectors. Therefore, partitioning the dataset of $N$ atoms into $K$ clusters can be done by dividing the rows of the descriptor matrix into $K$ similarity subsets. One can use a clustering method such as $k$-means clustering algorithm to partition $N$ vectors in $M$ dimensions into $K$ separate clusters. \revise{However, clustering in very high dimensions can be very expensive, since $M$ is typically large.}

To reduce computational cost, we consider a dimensionality reduction method to compress the descriptor matrix into a lower-dimensional matrix of size $N$ by $J$, where $J$ is considerably less than $M$. In this paper, principal component analysis is used to obtain the low-dimensional descriptor matrix as follows
\begin{equation}
\bm{\mathcal{B}} = \bm{\mathcal{D}} \ \bm W \ .    
\end{equation}
Here $\bm W \in \mathbb{R}^{M \times J}$ consists of the first $J$ eigenvectors of the eigenvalue decomposition $\bm{\mathcal{D}}^T \bm{\mathcal{D}} = \bm W \bm \Lambda \bm W^T$, and the eigenvalues are ordered from the largest to the smallest.

For multi-element systems, the clustering method is applied to the local descriptor matrix for each element as follows. The matrix  $\bm{\mathcal D}$ is split into $\bm{\mathcal D}^{e}, 1 \le e \le N_e$, where $\bm{\mathcal D}^{e}$ is formed by gathering the rows of $\bm{\mathcal D}$ for all atoms of the element $e$. For each element $e$, we compute 
\begin{equation}
\bm{\mathcal{B}}^{e} = \bm{\mathcal{D}}^{e} \ \bm W^{e}    
\end{equation}
where $\bm W^{e}$ consists of the $J$ eigenvectors of the eigenvalue decomposition $\left( \bm{\mathcal{D}}^{e} \right)^T \bm{\mathcal{D}}^{e} = \bm W^{e} \bm \Lambda \left( \bm W^{e} \right)^T$.  

\begin{figure*}
\includegraphics[width=\textwidth]{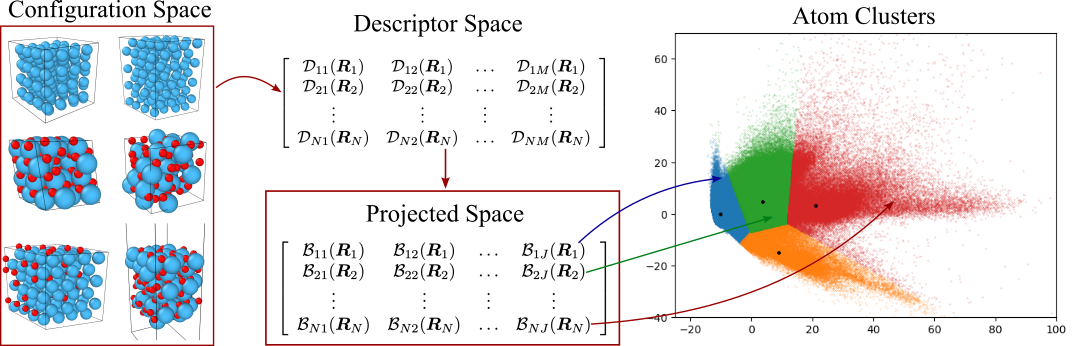}
\caption{\label{figcluster} Dataset of atom configurations is partitioned into atom clusters by using principal component analysis and the $k$-means algorithm.}
\end{figure*}

Next, we apply $k$-means clustering method to partition the rows of the matrix $\bm{\mathcal{B}}^{e}$ into $K$ separate clusters. We denote the centroids of the clusters by $\bm{\mathcal{C}}^{e}_k, 1 \le k \le K, 1 \le e \le N_e$. The clustering scheme allows us to divide the diverse dataset into smaller subsets, each characterized by similar data points which share the common attributes. This approach captures the diversity inherent in the dataset by identifying distinct atomic environments within the dataset.
Figure \ref{figcluster} illustrates the process of partitioning the dataset into several atom clusters. 

While local descriptors are used to partition $N$ atoms of the data set into $K$ subsets of atoms for each element, global descriptors can be used for partitioning the configurations of the dataset into $K$ subsets of configurations. For the dataset of $N_{\rm config}$ configurations, we sum the relevant rows of the local descriptor matrix $\bm{\mathcal D}$ to obtain the global descriptor matrix $\bm{\mathcal G}$ of size $N_{\rm config}$ by $M$. We then apply the above procedure to $\bm{\mathcal{G}}$ to obtain the desired configuration subsets. By training potentials on these subsets separately, we can construct MLIPs that are tailored to specific atomic environments.

\subsection{Many-body many-potential expansion}

We begin by introducing the atomic energies associated with the partitioned subsets
\begin{equation}
\label{eq29w}
 \mathcal{E}_{ik}(\bm R, \bm Z) = \sum_{m=1}^M c_{mk} \mathcal{D}_{im}  (\bm R, \bm Z), \quad 1 \le k \le K , 
\end{equation}
where the coefficients $c_{mk}$ are fitted against the QM data. For simplicity of exposition, the same local descriptors are used to define the atomic energies, although we allow for different local descriptors to be used for each subset. To construct a single potential energy surface, we introduce a many-body many-potential expansion
\begin{equation}
\label{eq30w}
E_i(\bm R, \bm Z) = \sum_{k=1}^K  \mathcal{P}_{ik}(\bm R, \bm Z)  \mathcal{E}_{ik}(\bm R, \bm Z) 
\end{equation}
where 
\begin{equation}
\label{eq31w}
 \sum_{k=1}^K  \mathcal{P}_{ik}(\bm R, \bm Z)  = 1, \quad \mathcal{P}_{ik}(\bm R, \bm Z)  \ge 0 .
\end{equation}
Thus, the atomic energy at atom $i$ is a weighted sum of the individual contributions from the $K$ subsets. Note that $\mathcal{P}_{ik}$ denotes the probability of atom $i$ belonging to the $k$th subset. Like the local descriptors, the probabilities depend on the local atomic environment of the central atom $i$.

By inserting (\ref{eq29w}) into the many-potential expansion (\ref{eq30w}) and summing over index $i$, we obtain the PES as
\begin{equation}
\label{eq32wa}
E_{\rm T}(\bm R, \bm Z) = \sum_{i=1}^{N_a} \sum_{k=1}^K \sum_{m=1}^M c_{mk}  \mathcal{P}_{ik}(\bm R, \bm Z)  \mathcal{D}_{im}  (\bm R, \bm Z) .
\end{equation}
The quantities $\mathcal{Q}_{ikm}(\bm R, \bm Z) = \mathcal{P}_{ik}(\bm R, \bm Z)  \mathcal{D}_{im}  (\bm R, \bm Z) $ shall be called {\em environment-adaptive descriptors}. Hence, we can write the PES as follows
\begin{equation}
\label{eq32w}
E_{\rm T}(\bm R, \bm Z) = \sum_{i=1}^{N_a} \sum_{l=1}^{KM} c_l  \mathcal{Q}_{il}(\bm R, \bm Z) ,
\end{equation}
where $l$ is a linear indexing of $k$ and $m$. The coefficients $c_l$ are sought as solution of a least squares
problem that minimizes a loss function defined as the weighted mean squared errors between the predicted energies/forces and  the QM energies/forces for all configurations in the training dataset.

For the single cluster case $K = 1$, the EA model (\ref{eq32wa}) reduces to the standard linear model
\begin{equation}
\label{eq33wa}
E_{\rm T}(\bm R, \bm Z) = \sum_{i=1}^{N_a} \sum_{m=1}^M c_{m}  \mathcal{D}_{im}  (\bm R, \bm Z) .
\end{equation}
\revise{Hence, the EA model (\ref{eq32w}) has $K$ times more descriptors and coefficients than  the standard linear model (\ref{eq33wa}). Because the size of the EA model increases with $K$, it can describe diverse environments in the dataset better than the standard linear model. Nonetheless, it is necessary to assess the EA model while varying $K$ and compare it against the standard linear model.}

\subsection{The probability functions}

It remains to calculate the probabilities $\mathcal{P}_{ik}$. They are defined in terms of the local descriptors as follows. First, we compute the low-dimensional descriptors. 
\begin{equation}
{\mathcal{B}}_{ij}(\bm R, \bm Z) = \sum_{m=1}^M W_{mj}^{Z_i} {\mathcal{D}}_{im} (\bm R, \bm Z), \quad 1 \le j \le J .  
\end{equation}
where $W_{mj}^e, 1 \le e \le N_e,$ are the PCA matrices. Next, we calculate the inverse of the square of the distance from the $k$th centroid as
\begin{equation}
{\mathcal{S}}_{ik}(\bm R, \bm Z) = \frac{1}{\sum_{j=1}^J \left({\mathcal{B}}_{ij}(\bm R, \bm Z) - \mathcal{C}_{kj}^{Z_i} \right)^2 } .
\end{equation}
Recall that $\mathcal{C}_{kj}^e$ are the centroids obtained from partitioning the dataset. Finally, the probabilities are calculated as 
\begin{equation}
{\mathcal{P}}_{ik}(\bm R, \bm Z) = \frac{{\mathcal{S}}_{ik}(\bm R, \bm Z)}{\sum_{l=1}^K {\mathcal{S}}_{il}(\bm R, \bm Z) }, \quad 1 \le k \le K .
\end{equation}
Hence, the probabilities are high when the distances between the low-dimensional descriptor vector and the centroids are small. The additional cost of evaluating the probabilities is only $O(JM + JK)$ per atom. This cost can be neglected, as it is considerably less than the cost of computing the local descriptors.

\subsection{Force calculation}
\label{forcecal}

Forces on atoms are calculated by differentiating the PES (\ref{eq32wa}) with respect to atom positions. To this end, we first compute the partial derivatives of the probabilities with respect to the local descriptors as
\begin{equation}
\label{eq36w}
\frac{\partial {\mathcal{P}}_{ik}}{\partial \mathcal D_{im}} = \frac{\partial {\mathcal{P}}_{ik}}{\partial \mathcal S_{il}}  \frac{\partial {\mathcal{S}}_{il}}{\partial \mathcal B_{ij}}  \frac{\partial {\mathcal{B}}_{ij}}{\partial \mathcal D_{im}}  .
\end{equation}
Here the Einstein summation convention is used to indicate the implicit summation over repeated indices except for the index $i$. The cost of evaluating the terms in (\ref{eq36w}) is $O(K^2 M  J)$ per atom. Next, we note that
\begin{equation}
\label{eq37w}
\frac{\partial {\mathcal{P}}_{ik}(\bm R, \bm Z)}{\partial \bm R} = \frac{\partial {\mathcal{P}}_{ik}}{\partial \mathcal D_{im}}  \frac{\partial {\mathcal{D}}_{im}}{\partial \bm R}   .
\end{equation}
Differentiating the PES (\ref{eq32wa}) with respect to atom positions yields 
\begin{equation}
\label{eq38w}
\frac{\partial E_{\rm T}(\bm R, \bm Z)}{\partial \bm R} =  c_{nk} \mathcal{D}_{in}    \frac{\partial {\mathcal{P}}_{ik}}{\partial \bm R} +    c_{ml} \mathcal{P}_{il}  \frac{\partial {\mathcal{D}}_{im}}{\partial \bm R}  .  
\end{equation}
By inserting (\ref{eq37w}) into (\ref{eq38w}), we obtain 
\begin{equation}
\frac{\partial E_{\rm T}(\bm R, \bm Z)}{\partial \bm R} =\left(  c_{nk}  \mathcal{D}_{in}   \frac{\partial {\mathcal{P}}_{ik}}{\partial \mathcal D_{im}}     +   c_{ml}   \mathcal{P}_{il}  \right) \frac{\partial {\mathcal{D}}_{im}}{\partial \bm R}  .
\end{equation}
The additional cost of evaluating the terms in the parenthesis is only $O(3MK)$ per atom. 

\revise{In summary, the additional cost of evaluating the forces on atoms is $O(K^2 M J)$ per atom for any $K > 1$. Since this cost is independent of the number of neighbors and linear in the number of local descriptors, it can be much smaller than the cost of computing the local descriptors and their derivatives with respect to atom positions. We can thus expect that the EA potentials are almost as fast as the standard linear potential. Therefore, the proposed method enhances the EA potentials without increasing computational cost.}

\section{Results and Discussions}
\label{results}

The EA potentials will be demonstrated and compared with the standard linear potential for Tantalum element and Indium Phosphide compound. For all potentials, the hyperparameters are fixed to $\alpha = 100, \beta = 1, \gamma = 10^{-12}$. In order to assess their performance, all potentials are trained on the same training datasets and validated on the same test datasets. We evaluate the potentials using the mean absolute errors (MAEs) of the predicted energies and forces
\begin{equation}
\begin{split}
\varepsilon_E & = \frac{1}{N_{\rm config}} \sum_{n = 1}^{N_{\rm config}} |E_{n} - E_n^{\rm DFT}| \\
\varepsilon_F & = \frac{1}{N_{\rm force}} \sum_{n = 1}^{N_{\rm force}} |F_{n} - F_{n}^{\rm DFT}| \\
\end{split}
\end{equation}
where $N_{\rm config}$ is the number of configurations in a data set, and $N_{\rm force}$ is the total number of force components for all the configurations in the same data set. 
Both the source code and the data are available upon request to facilitate the reproduction of our work.

\subsection{Results for Tantalum}

The Ta data set contains a wide range of configurations to adequately sample the important regions of the potential energy surface \cite{Thompson2015}. The data set includes 12 different groups such as surfaces, bulk structures, defects, elastics for BCC, FCC, and A15 crystal structures, and high temperature liquid. The database was used to create a SNAP potential \cite{Thompson2015} which successfully describes a wide range of properties such as energetic properties of solid tantalum phases, the size and shape of the Peierls barrier for screw dislocation motion in BCC tantalum, as well as both the structure of molten tantalum and its melting point.  We train eight EAML models on the Ta dataset for different values of $M$ and $K$. Table \ref{tabta1} shows the number of descriptors for the eight EAML potentials.  Note that all potentials have a one-body descriptor to account for isolated energies. The inner and outer cut-off distances are set to $r_{\rm in} = 1.0$\AA \ and $r_{\rm cut} = 5.0$\AA, respectively. Furthermore, we use $J = 2$ in all cases.

\begin{table}[htbp]
	\caption{The number of radial basis functions $N_r$, number of angular basis functions $K_a$, number of descriptors $N_d$ for two-body, three-body, and four-body POD descriptors.}
	\label{tabta1}
	\begin{tabular}{|c|cc|ccc|ccc|c|}
		\cline{1-10}
		\textbf{Ta} &
		 \multicolumn{2}{|c|}{\textbf{two-body}} & \multicolumn{3}{c|}{\textbf{three-body}} & 
		 \multicolumn{3}{c|}{\textbf{four-body}} & \textbf{All} \\
		\hline
		Case & $N_r$ & $N_d$ & $N_r$ & $K_a$ & $N_d$ & $N_r$ & $K_a$ & $N_d$ & $M$  \\
		\hline
1  &  4  &  4  &  2  &  2  &  4  &  0  & 0 & 0  & 8   \\  
2  &  5  &  5  &  3  &  3  &  9  &  0  & 0 & 0  & 14   \\  
3  &  6  &  6  &  4  &  4  &  16  &  0  & 0 & 0  & 22   \\  
4  &  7  &  7  &  5  &  5  &  25  &  0  & 0 & 0  & 32   \\  
5  &  8  &  8  &  6  &  5  &  30  &  3  & 2 & 6  & 44   \\  
6  &  9  &  9  &  7  &  5  &  35  &  4  & 4 & 16  & 60   \\  
7  &  10  &  10  &  8  &  6  &  48  &  5  & 4 & 20  & 78   \\  
8  &  11  &  11  &  9  &  6  &  54  &  5  & 7 & 35  & 100   \\  
		\hline
	\end{tabular}
\end{table}

\begin{table}[htbp]
\caption{
Training errors in energies and forces for EAML potentials are listed in Table \ref{tabta1}. The units for the MAEs in energies and forces are meV/atom and meV/\AA, respectively.
}
\label{tabta2}
\begin{ruledtabular}
\begin{tabular}{c|cc|cc|cc|cc}
&
\multicolumn{2}{c|}{$K=1$} &
\multicolumn{2}{c|}{$K=2$} &
\multicolumn{2}{c|}{$K=3$} &
\multicolumn{2}{c}{$K=4$} \\   
\hline
Case & $\varepsilon_E$ & $\varepsilon_F$ & $\varepsilon_E$ & $\varepsilon_F$ & $\varepsilon_E$ & $\varepsilon_F$ & $\varepsilon_E$ & $\varepsilon_F$  \\
\hline 
 1  &  74.6  &  132.65  &  35.02  &  161.45  &  25.47  &  178.24  &  20.01  &  179.69  \\  
 2  &  55.37  &  237.93  &  22.98  &  228.11  &  14.43  &  208.26  &  6.72  &  130.22  \\  
 3  &  37.63  &  209.94  &  8.60  &  128.45  &  5.98  &  99.94  &  4.94  &  101.05  \\  
 4  &  10.47  &  106.69  &  4.43  &  95.77  &  2.51  &  70.99  &  2.00  &  67.16  \\  
 5  &  8.02  &  96.94  &  3.14  &  74.04  &  1.75  &  62.49  &  1.49  &  59.05  \\  
 6  &  4.00  &  88.38  &  1.58  &  62.49  &  1.19  &  52.46  &  1.02  &  50.82  \\  
 7  &  2.37  &  76.73  &  1.20  &  55.82  &  0.80  &  49.94  &  0.60  &  45.51  \\  
 8  &  1.77  &  64.11  &  0.83  &  48.91  &  0.56  &  43.10  &  0.49  &  39.56  \\  
\end{tabular}
\end{ruledtabular}
\end{table}

\revise{Table \ref{tabta2} displays training errors in energies and forces predicted by EAML potentials for different values of the number of the descriptors listed in Table \ref{tabta1} and for
$K = 1, 2, 3, 4$. We see that both the energy and force errors decrease as the number of descriptors increases. As $M$ increases from $8$ to 100, the energy errors drop by a a factor of 20, while the force errors drop by a factor of 4. As $K$ increases from $1$ to $4$,  the energy errors decrease by a factor of 4, while the force errors decrease  by a factor of 1.5.  The energy errors reach 1.77 meV/atom, 0.83 meV/atom, 0.56 meV/atom, and 0.48 meV/atom for $K=1,2,3,$ and 4, respectively. These energy errors are below the typical numerical errors of DFT calculations. The force errors reach  64.11 meV/\AA, 48.91 meV/\AA,  43.10 meV/\AA, and 39.56 meV/\AA. These force errors are acceptable for most applications.  The errors decrease quite rapidly as $K$ increases from 1 to 2. However, as $K$ increases from $2$ to $4$, the rate of error decrease slows down considerably for this dataset and we observe a smaller improvement. Hence, we may only gain marginal improvements by increasing $K$ beyond 4.}

\revise{Table \ref{tabta3} provides the training errors in energy and forces for each of the 12 groups for $M=60$. The force errors for Bulk A15, Bulk BCC, and Bulk FCC are zero, because their  structures are at equilibrium states and thus have zero atomic forces. The Surface group tends to have high energy errors than other groups, while the Liquid group has the highest force errors. The liquid structures depend strongly on the repulsive interactions that occur when two atoms approach each other. Consequently, it is more difficult to predict atomic forces of the liquid phase since the liquid configurations are very different from those of the equilibrium solid crystals.  It is also more difficult to predict energies of surface configurations because the surfaces of BBC crystals tend to be rather open with surface atoms exhibiting rather low coordination numbers.}

\begin{table}[htbp]
\caption{\label{tabta3}%
Energy and force errors for EAML potentials with $M = 60$ for different configuration groups. The units for the MAEs in energies and forces are meV/atom and meV/\AA, respectively. 
}
\begin{ruledtabular}
\begin{tabular}{c|cc|cc|cc|cc}
&
\multicolumn{2}{c|}{$K=1$} &
\multicolumn{2}{c|}{$K=2$} &
\multicolumn{2}{c|}{$K=3$} &
\multicolumn{2}{c}{$K=4$} \\   
\hline
Group & $\varepsilon_E$ & $\varepsilon_F$ & $\varepsilon_E$ & $\varepsilon_F$ & $\varepsilon_E$ & $\varepsilon_F$ & $\varepsilon_E$ & $\varepsilon_F$  \\
\hline 
Disp. A15 & 1.94  &  125.28  &  0.42  &  80.82  &  1.66  &  76.02  &  1.35  &  71.27  \\  
Disp. BCC & 11.71  &  140.57  &  5.81  &  110.24  &  5.47  &  92.13  &  5.13  &  89.89  \\  
Disp. FCC & 1.79  &  106.22  &  3.36  &  77.38  &  2.95  &  51.53  &  2.84  &  51.88  \\  
Elas. BCC & 0.91  &  0.04  &  0.55  &  0.01  &  0.38  &  0.01  &  0.38  &  0.00  \\  
Elas. FCC & 0.72  &  0.16  &  0.47  &  0.13  &  0.38  &  0.13  &  0.34  &  0.12  \\  
GSF 110 & 3.78  &  41.35  &  2.03  &  15.84  &  1.53  &  17.09  &  1.26  &  15.8  \\  
GSF 112 & 5.43  &  59.10  &  3.27  &  51.41  &  2.30  &  42.01  &  2.12  &  41.55  \\  
Liquid & 11.28  &  371.38  &  2.40  &  262.04  &  2.12  &   223.8  &  2.81  &  216.9  \\  
Surface & 13.66  &  62.00  &  5.66  &  40.02  &  4.67  &  28.97  &  3.60  &  28.06  \\  
Bulk A15 & 4.87  &  0  &  2.19  &  0  &  1.32  &  0  &  1.16  &  0  \\  
Bulk BCC & 11.96  &  0  &  3.47  &  0  &  2.32  &  0  &  1.49  &  0  \\  
Bulk FCC & 13.59  &  0  &  2.74  &  0  &  1.69  &  0  &  1.31  &  0  \\  
\end{tabular}
\end{ruledtabular}
\end{table}




\revise{Next, we investigate the influence of training datasets on model performance. To this end, we train four potentials on different training datasets with $M=14$. The first three potentials are standard linear models, while the fourth potential is an EAML potential with $K=2$ clusters. Table~\ref{tabta4} displays the MAEs in energies for the four potentials.  The first potential, trained exclusively on the Bulk A15 group, demonstrates a small error of 2.54 meV/atom for this group but a very large error of 1070.7 meV/atom for the Bulk FCC group. Conversely, the second potential, trained on the Bulk FCC group, shows a small error of 11.41 meV/atom for its training group and a significant error of 218.57 meV/atom for the Bulk A15 group. While each potential performs well on the dataset it was trained on, its predictions for the other group are highly inaccurate. The third potential, trained on both the Bulk A15 and Bulk FCC groups, exhibits more balanced errors of 37.63 meV/atom and 41.57 meV/atom for the Bulk A15 and Bulk FCC groups, respectively. The fourth potential, trained on both groups with $K=2$ clusters, achieves superior accuracy with errors of 2.06 meV/atom and 3.88 meV/atom for the Bulk A15 and Bulk FCC groups, respectively. These results underscore the importance of diverse training datasets and demonstrate the substantial improvement of the EAML model over the standard linear model.}


\begin{table}[htbp]
\caption{\label{tabta4}%
Energy errors for the standard linear potentials and EAML potential with $M = 14$ when they are trained on different training groups and validated on Bulk A15 and Bulk FCC groups. The unit for the MAEs in energies is meV/atom.
}
\begin{tabular}{|c|c|clcl|}
\hline
\multirow{2}{*}{Training data} & \multirow{2}{*}{Clusters} & \multicolumn{4}{c|}{Test Groups}                                   \\ \cline{3-6} 
                           &                           & \multicolumn{2}{c|}{Bulk A15} & \multicolumn{2}{c|}{Bulk FCC} \\ \hline
Bulk A15                          & 1                         & \multicolumn{2}{c|}{2.54}     & \multicolumn{2}{c|}{1070.7}   \\ 
Bulk FCC                          & 1                         & \multicolumn{2}{c|}{218.57}   & \multicolumn{2}{c|}{11.41}    \\
Bulk A15 \& Bulk  FCC                          & 1                         & \multicolumn{2}{c|}{37.63}    & \multicolumn{2}{c|}{41.57}    \\ \hline 
Bulk A15 \& Bulk FCC                          & 2                         & \multicolumn{2}{c|}{2.06}     & \multicolumn{2}{c|}{3.88}     \\ \hline
\end{tabular}
\end{table}

\begin{figure*}
\includegraphics[width=0.9\textwidth]{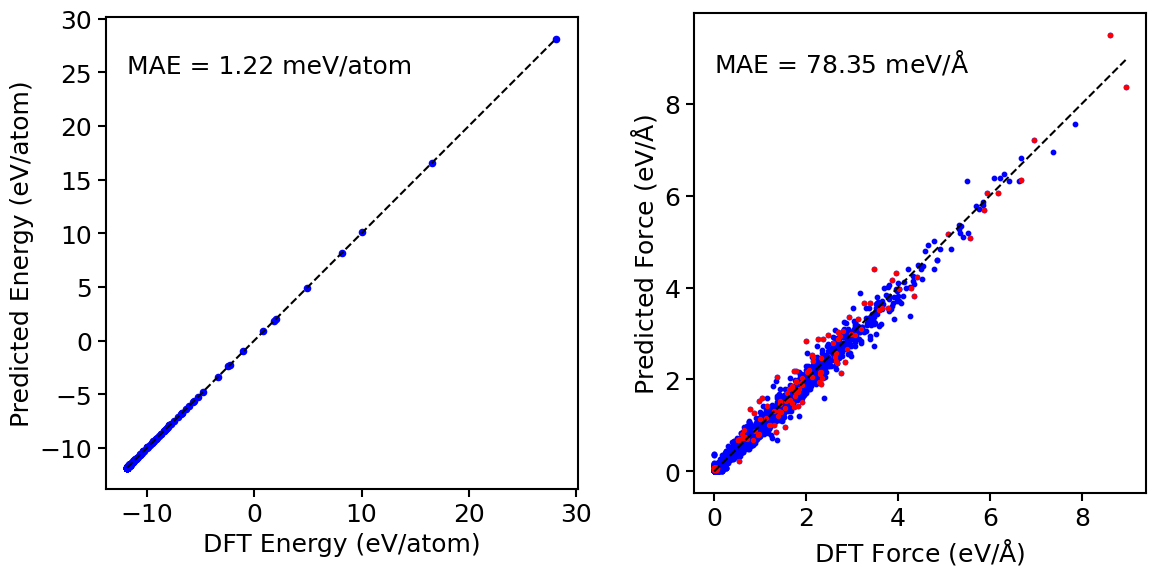}
\caption{\label{figparityTa} Energy parity (left) and atomic force parity (right) plots for Ta for Case 5 with $K=4$. For force parity, atoms that do not belong to one specific the environment with $\displaystyle{\max_{k}\mathcal{P}_k<0.7}$ (red dots) have similar errors as atoms that belong to one specific environment with $\displaystyle{\max_{k}\mathcal{P}_k \ge 0.7}$ (blue dots).}
\end{figure*}




\revise{Figure~\ref{figparityTa} shows the energy and atomic force parity plots for Ta for $M=44$ with $K=4$ environments from Table~\ref{tabta1}.  We note that atoms with $\displaystyle{\max_{k}\mathcal{P}_k<0.7}$ that do not belong exclusively to one environment have similar force errors as the other atoms. This shows the ability of the EAML model to capture atomic environments that are a mixture of several distinct environments, thereby making itself more transferable than the standard linear model. The transferability of the EAML model can be attributed to several factors. First, the EAML model has more capacity than the standard linear potential because it has $K$ times more trainable coefficients. Second, owing to the probability functions that vary with the neighborhood of the central atom, the EAML model adapts itself according to the local atomic environments to capture atomic interactions more accurately than the linear model. Third, the products of the probability functions and the descriptors contain higher body interactions than the descriptors themselves, rendering the EAML model higher body order than the linear model.}

\begin{figure*}
\begin{subfigure}[b]{0.32\textwidth}
\centering
\includegraphics[width=\textwidth]{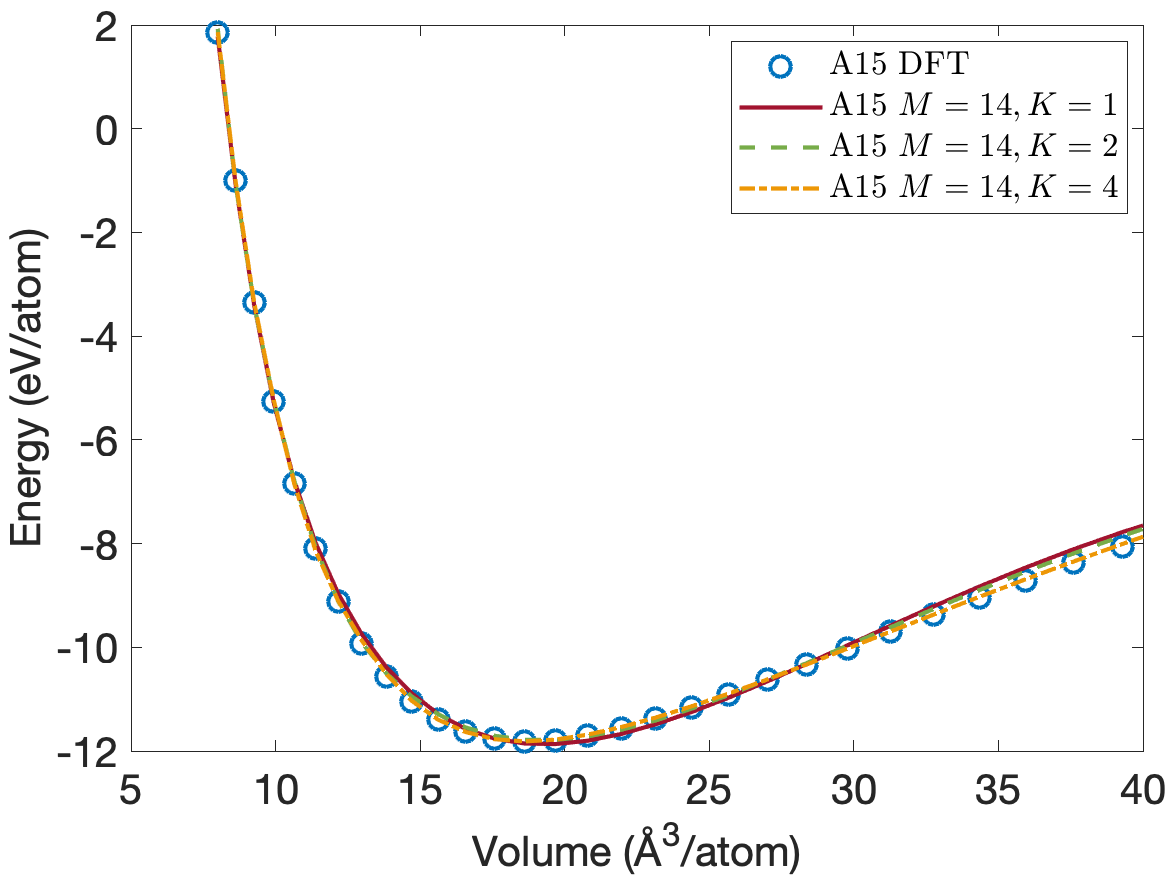}
\caption{A15}
\end{subfigure}
\begin{subfigure}[b]{0.32\textwidth}
\centering
\includegraphics[width=\textwidth]{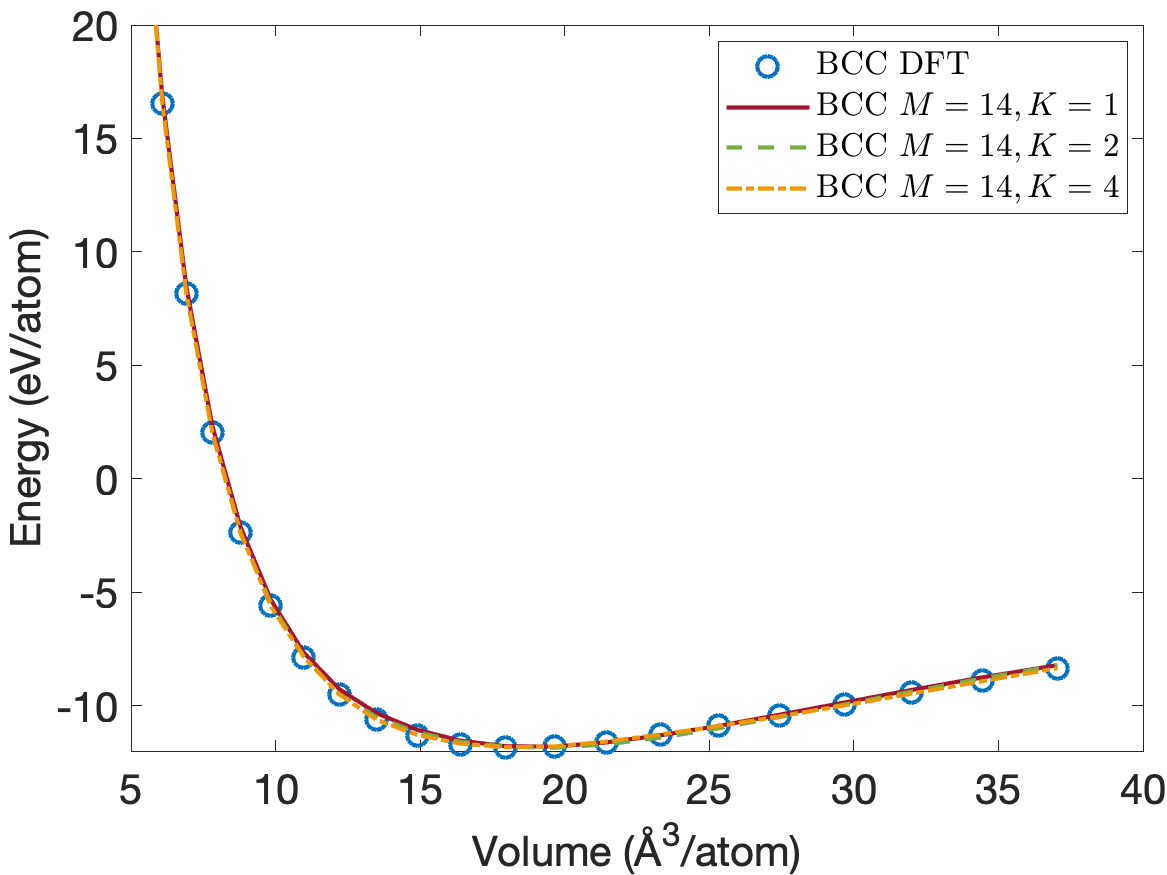}
\caption{BCC}
\end{subfigure}
\begin{subfigure}[b]{0.32\textwidth}
\centering
\includegraphics[width=\textwidth]{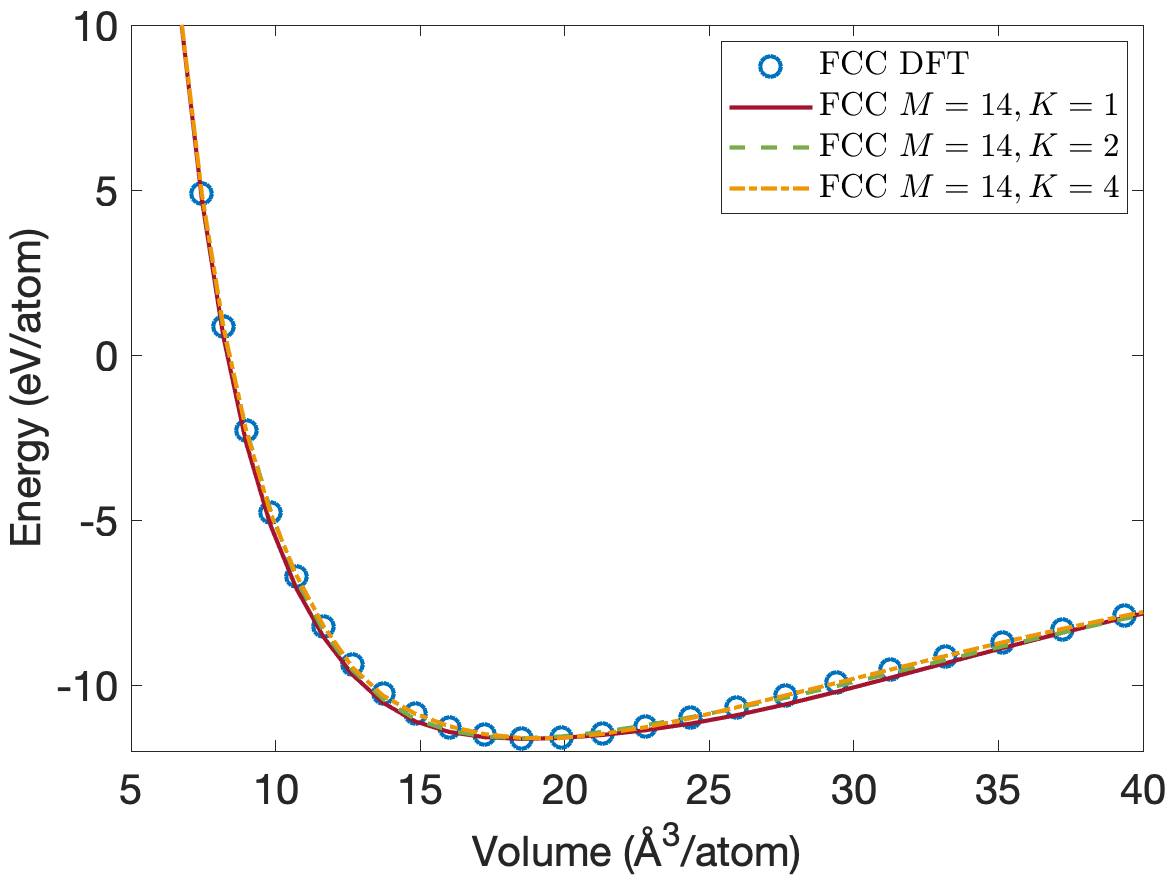}
\caption{FCC}
\end{subfigure}
\caption{\label{figta1} Energy per atom versus volume per atom for A15, BCC, and FCC crystal structures for EAML potentials using $M=14$ descriptors in comparison with DFT data.}
\end{figure*}

\begin{figure*}
\begin{subfigure}[b]{0.32\textwidth}
\centering
\includegraphics[width=\textwidth]{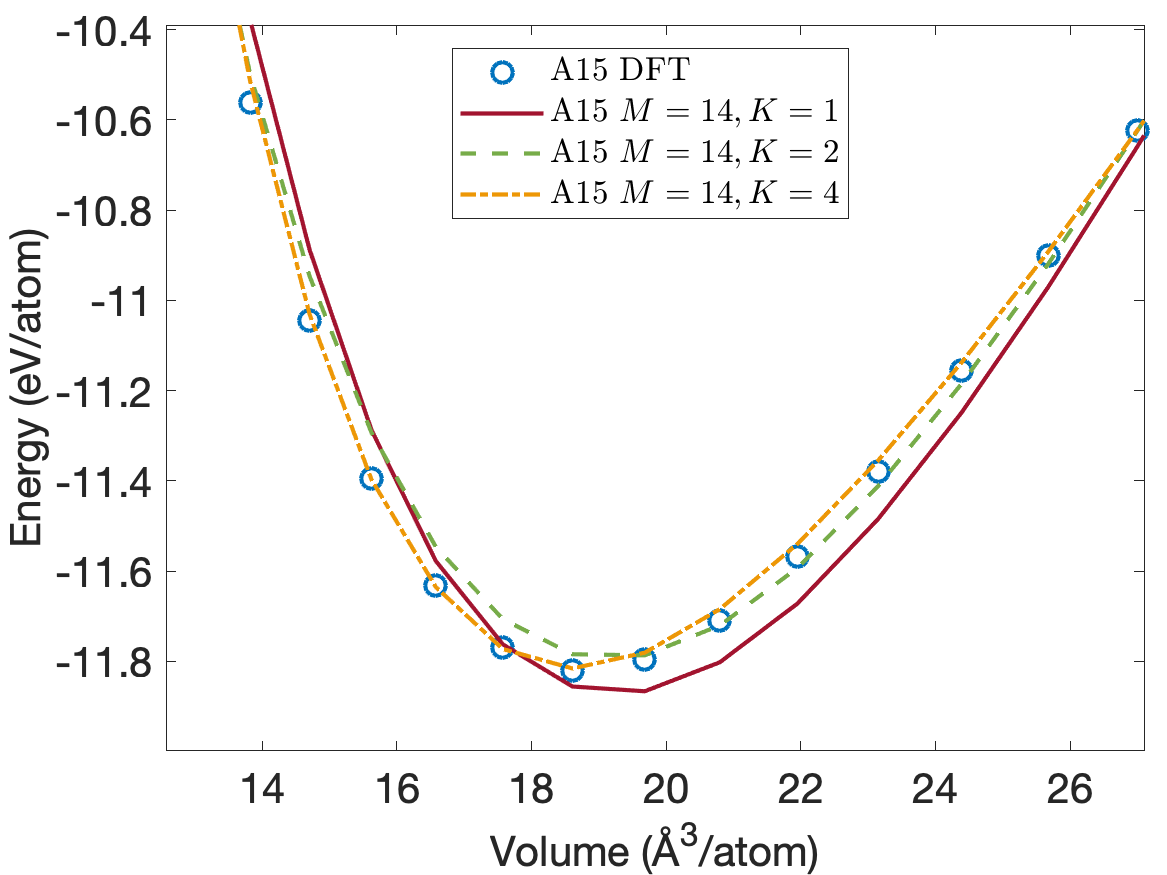}
\caption{A15}
\end{subfigure}
\begin{subfigure}[b]{0.32\textwidth}
\centering
\includegraphics[width=\textwidth]{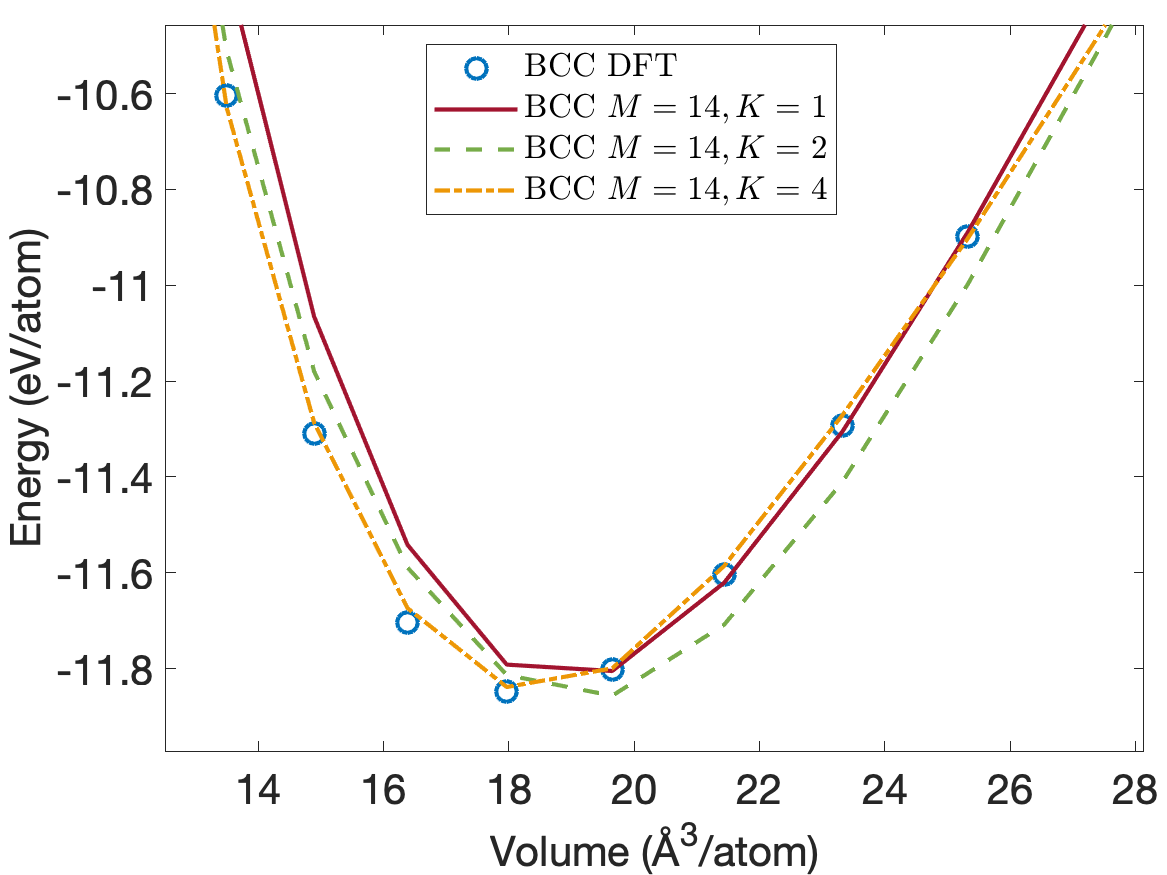}
\caption{BCC}
\end{subfigure}
\begin{subfigure}[b]{0.32\textwidth}
\centering
\includegraphics[width=\textwidth]{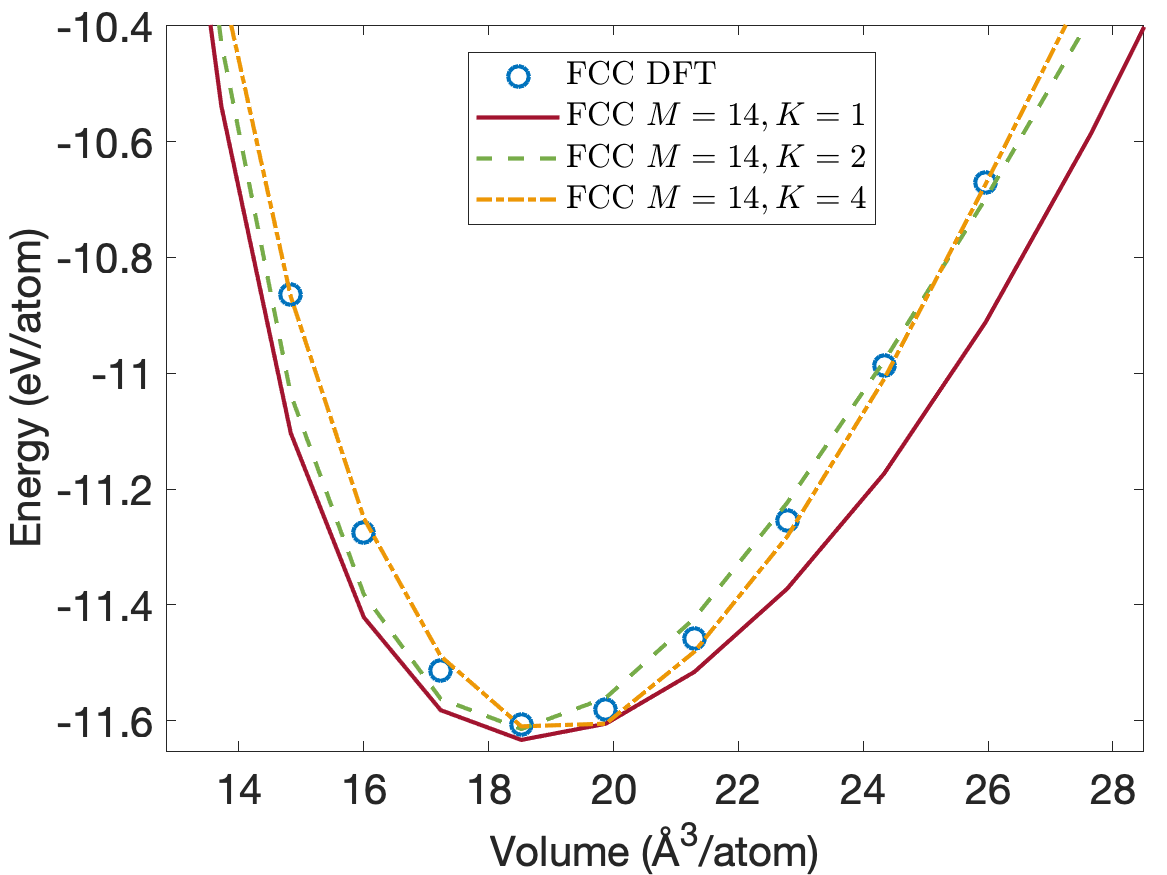}
\caption{FCC}
\end{subfigure}
\caption{\label{figta2} Close-up view near the minimum energy for the energy per atom versus volume per atom curves.}
\end{figure*}

\revise{Figure \ref{figta1} plots the energy per atom as a function of volume per atom for A15, BCC, FCC crystal structures. The FCC phase has a minimum energy about 0.2eV/atom above the BCC and A15 phases. We see that the energies are predicted accurately for the whole volume range with using only $M=14$ descriptors. Figure \ref{figta2} plots the close-up view of the energy curve near the minimum energy. The predicted energy curves for $K=4$ are almost indistinguishable from the DFT energy curves for BCC, FCC and A15 phases.} 

\begin{figure*}
\begin{subfigure}[b]{0.49\textwidth}
\centering
\includegraphics[width=\textwidth]{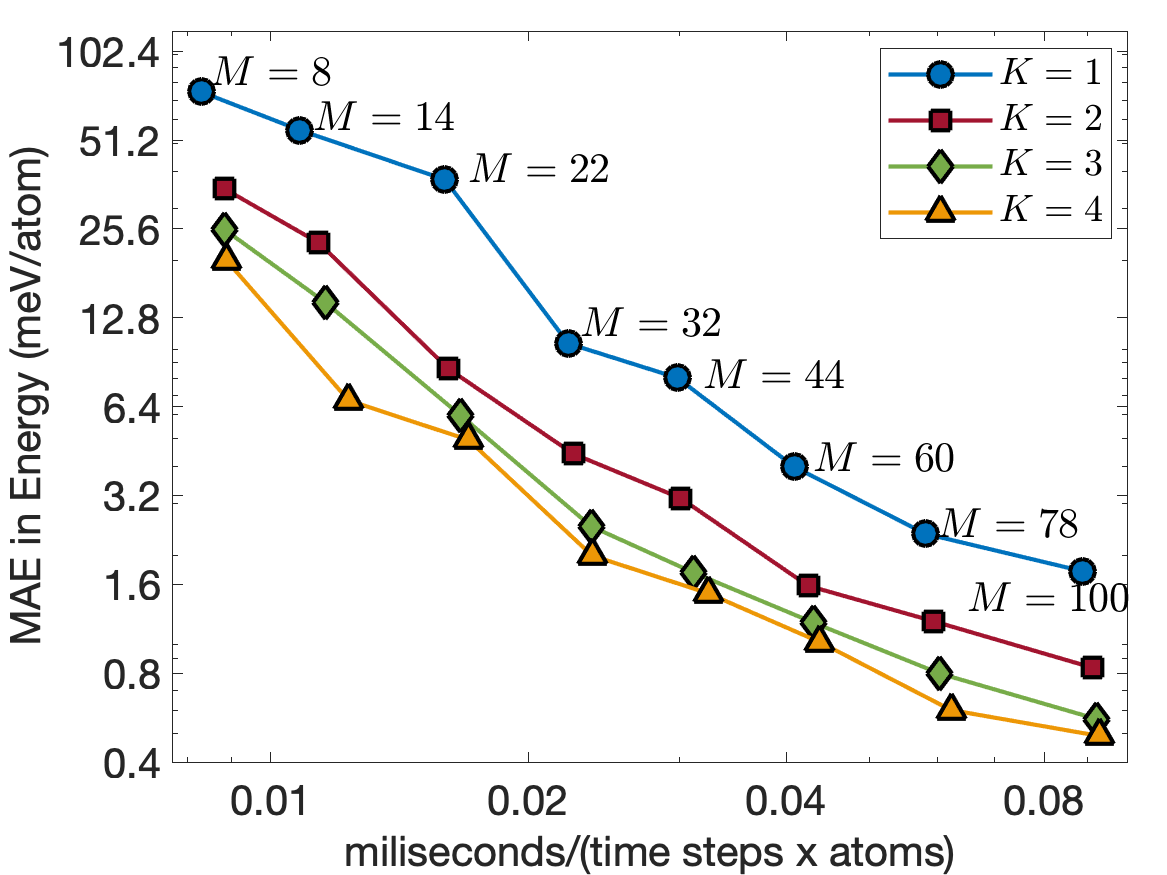}
\caption{Energy}
\end{subfigure}
\begin{subfigure}[b]{0.49\textwidth}
\centering
\includegraphics[width=\textwidth]{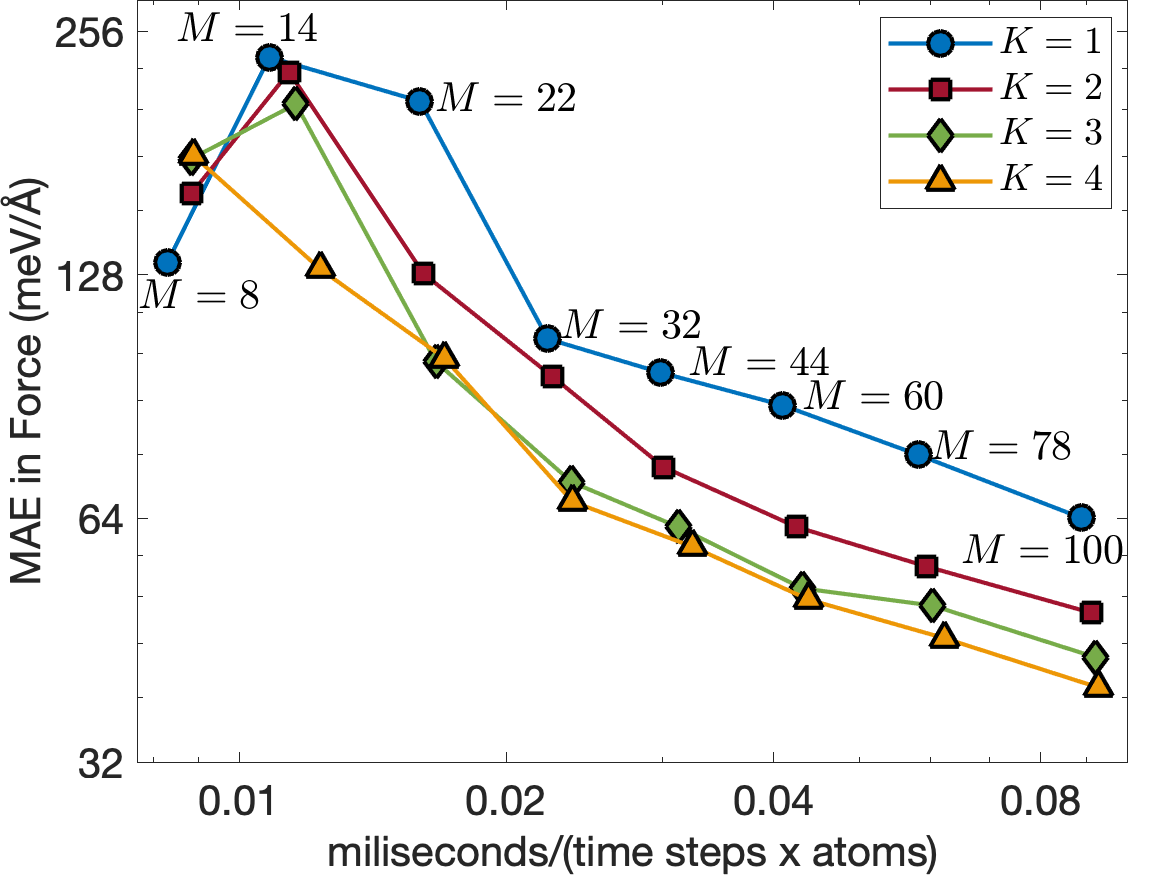}
\caption{Force}
\end{subfigure}
\caption{\label{figta3} Training errors versus the computational cost of MD simulations for the Ta system of 16000 atoms. MD simulations are performed using LAMMPS \cite{Thompson2022} on a CPU core of Intel i7-1068NG7 2.3 GHz for EAML potentials with different numbers of descriptors and numbers of clusters.}
\end{figure*}

\revise{Figure \ref{figta3}
illustrates the trade-off between computational cost and training error for $K=1,2,3,4$. The computational cost is measured in terms of milisecond per time step per atom for MD simulations. These MD simulations are performed using LAMMPS \cite{Thompson2022} on a CPU core of Intel i7-1068NG7 2.3 GHz with $20 \times 20 \times 20$ bulk supercell containing 16000 Tantalum atoms. The EAML potentials with $K > 1$ are almost as fast as the standard linear potential for the same number of descriptors, having the computational cost almost independent of $K$. This is consistent with the computational complexity analysis discussed in Subsection \ref{forcecal}.  Furthermore, the EAML potential with $M = 32, K = 4$ is more accurate and 3 times faster than the standard linear potential with $M = 100$. The results show the superior performance of the EAML potentials.}


\subsection{Results for Indium Phosphide}

The InP dataset contains a wide range of configurations to adequately sample the important regions of the potential energy surface. It was generated by Cusentino et. al \cite{Cusentino2020} using the Vienna Ab Initio Simulation Package to demonstrate the explicit multi-element SNAP  potential. The InP dataset also contains high-energy defects which are intended to study radiation damage effects where collision cascades of sufficiently high energy leave behind high formation energy point defects. Furthermore, the dataset includes configurations for uniform expansion and compression (Equation of State), random cell
shape modifications (Shear group), and uniaxially strained
(Strain group) unit cells for zincblende crystal structure. In total, the dataset has 1894 configurations with atom counts per configuration ranging from 8 to 216. The training set is 80\% of the InP dataset, while the entire InP dataset is used as the test set. The inner and outer cut-off distances are set to $r_{\rm in} = 0.8$\AA \ and $r_{\rm cut} = 5$\AA, respectively.



\revise{Table \ref{tab1b} displays the number of descriptors for six different cases.   Table \ref{table7} provides test errors in energies and forces for cases listed in Table \ref{tab1b} and for $K = 1, 2, 3, 4$. Both the energy and force errors decrease as $M$ and $K$ increases. As $M$ increases from $40$ to $370$, the energy errors drop more than a factor 10, while the force errors drop by a factor of 4. As $K$ increases from $1$ to $4$, the energy errors decrease by a factor of $4$ and the force errors decrease by a factor 2. The MAEs in energies reach 0.97 meV/atom for $K=1$, 0.42 meV/atom for $K=2$, 0.31 meV/atom for $K=3$, and 0.23 meV/atom for $K=4$. These energy errors are generally below the limits of DFT errors. The MAEs in forces reach  16.70, 11.16, 9.05, 7.79 meV/\AA \ for $K=1,2,3, 4$, respectively. These force errors are acceptable for most applications.}  


\begin{table}[htbp]
	\caption{The number of radial basis functions $N_r$, number of angular basis functions $K_a$, number of descriptors $N_d$ for two-body, three-body, and four-body POD descriptors. Note that $N_d = N_r N_e^2$ for two-body PODs, $N_d = N_rK_a N_e^2 (N_e+1)/2$ for three-body PODs, $N_d = N_rK_a N_e^2 (N_e+1)(N_e+2)/6$ for four-body PODs.}
	\label{tab1b}
	\begin{tabular}{|c|cc|ccc|ccc|c|}
		\cline{1-10}
		\textbf{InP} &
		 \multicolumn{2}{|c|}{\textbf{two-body}} & \multicolumn{3}{c|}{\textbf{three-body}} & 
		 \multicolumn{3}{c|}{\textbf{four-body}} & \textbf{All} \\
		\hline
		Case & $N_r$ & $N_d$ & $N_r$ & $K_a$ & $N_d$ & $N_r$ & $K_a$ & $N_d$ & $M$  \\
		\hline
1  &  4  &  16  &  2  &  2  &  24  &  0  & 0 & 0  & 40   \\  
2  &  5  &  20  &  3  &  3  &  54  &  0  & 0 & 0  & 74   \\  
3  &  6  &  24  &  4  &  4  &  96  &  0  & 0 & 0  & 120   \\  
4  &  7  &  28  &  5  &  5  &  150  &  0  & 0 & 0  & 178   \\  
5  &  8  &  32  &  6  &  5  &  180  &  3  & 2 & 48  & 260   \\  
6  &  9  &  36  &  7  &  5  &  210  &  4  & 4 & 128  & 374   \\  
		\hline
	\end{tabular}
\end{table}

\begin{table}[htbp]
\caption{\label{table7}%
Test errors in energies and forces for EAML potentials listed in Table \ref{tab1b}. The units for the MAEs in energies and forces are meV/atom and meV/\AA, respectively.
}
\begin{ruledtabular}
\begin{tabular}{c|cc|cc|cc|cc}
&
\multicolumn{2}{c|}{$K=1$} &
\multicolumn{2}{c|}{$K=2$} &
\multicolumn{2}{c|}{$K=3$} &
\multicolumn{2}{c}{$K=4$} \\   
\hline
Case & $\varepsilon_E$ & $\varepsilon_F$ & $\varepsilon_E$ & $\varepsilon_F$ & $\varepsilon_E$ & $\varepsilon_F$ & $\varepsilon_E$ & $\varepsilon_F$  \\
\hline 
 1  &  11.63  &  61.00  &  6.44  &  48.13  &  4.67  &  36.76  &  4.11  &  33.30  \\  
 2  &  4.97  &  38.71  &  3.49  &  30.27  &  2.52  &  24.00  &  2.01  &  22.05  \\  
 3  &  3.30  &  33.80  &  1.55  &  21.28  &  1.11  &  17.30  &  0.91  &  15.34  \\  
 4  &  2.67  &  27.26  &  1.20  &  17.81  &  0.68  &  14.56  &  0.46  &  12.74  \\  
 5  &  1.71  &  20.91  &  0.70  &  14.28  &  0.52  &  12.03  &  0.38  &  10.98  \\  
 6  &  0.97  &  16.70  &  0.43  &  11.10  &  0.28  &  8.91  &  0.20  &  7.79  \\  
\end{tabular}
\end{ruledtabular}
\end{table}

\revise{Table \ref{table8} provides the test errors in energy and forces for each of the 19 groups in the dataset for $M=178$. Point defects are created when atoms become vacant at lattice sites (vacancy defect),  occupy locations in the crystal structure at which there is usually no atom (interstitial defect), or exchange positions with other atoms of different types (antisite defect). The defect groups have higher errors than the other groups. The In$^{\rm S}_{\rm v}$ group has the highest mean absolute error in energies, while the  In$^{\rm S}_{\rm v}$P$^{\rm S}_{\rm v}$ group has the highest mean absolute error in forces. We see that increasing $K$ reduces the energy and force errors across all groups.}

\begin{table}[htbp]
\caption{\label{table8}%
Energy and force errors for EAML potentials with $M = 178$ for different configuration groups. The units for the MAEs in energies and forces are meV/atom and meV/\AA, respectively. Point defects are created from an equilibrium configuration by inserting atoms  (interstitial), removing atoms (vacancy), or exchanging atoms of different types (antisite). Subscripts correspond to vacancy(v), interstitial(i) and antisite(a). Superscripts correspond to large configurations of 216 atoms (L) and small configurations of 64 atoms (S). 
}
\begin{ruledtabular}
\begin{tabular}{c|cc|cc|cc|cc}
&
\multicolumn{2}{c|}{$K=1$} &
\multicolumn{2}{c|}{$K=2$} &
\multicolumn{2}{c|}{$K=3$} &
\multicolumn{2}{c}{$K=4$} \\   
\hline
Group & $\varepsilon_E$ & $\varepsilon_F$ & $\varepsilon_E$ & $\varepsilon_F$ & $\varepsilon_E$ & $\varepsilon_F$ & $\varepsilon_E$ & $\varepsilon_F$  \\
\hline 
Bulk & 5.79  &  0.00  &  2.26  &  0.00  &  1.50  &  0.00  &  0.91  &  0.00  \\  
EOS & 2.60  &  0.70  &  0.96  &  0.83  &  0.80  &  0.89  &  0.49  &  0.42  \\  
Shear & 0.26  &  25.56  &  0.50  &  8.45  &  0.17  &  5.23  &  0.11  &  3.64  \\  
Strain & 2.53  &  0.02  &  1.24  &  0.02  &  0.90  &  0.03  &  0.91  &  0.05  \\  
In$^{\rm L}_{\rm a}$ & 3.05  &  10.77  &  1.05  &  7.49  &  0.55  &  6.30  &  0.23  &  6.48  \\  
P$^{\rm L}_{\rm a}$ & 4.09  &  29.6  &  1.30  &  18.22  &  1.03  &  14.32  &  0.72  &  12.10  \\  
In$^{\rm L}_{\rm a}$P$^{\rm L}_{\rm a}$ & 6.68  &  23.23  &  3.00  &  16.83  &  1.71  &  13.96  &  1.14  &  13.52  \\  
In$^{\rm L}_{\rm i}$ & 4.68  &  19.82  &  1.82  &  13.80  &  1.13  &  11.36  &  0.73  &  10.35  \\  
P$^{\rm L}_{\rm i}$ & 3.34  &  15.68  &  1.28  &  9.96  &  0.87  &  8.55  &  0.61  &  7.69  \\  
P$^{\rm L}_{\rm v}$ & 3.65  &  7.95  &  0.73  &  5.17  &  0.22  &  5.62  &  0.09  &  5.09  \\  
In$^{\rm L}_{\rm v}$P$^{\rm L}_{\rm v}$ & 4.20  &  27.21  &  1.85  &  21.76  &  1.18  &  18.91  &  0.69  &  17.38  \\  
In$^{\rm S}_{\rm a}$ &  3.66  &  15.54  &  0.67  &  10.29  &  0.36  &  7.82  &  0.22  &  7.31  \\  
P$^{\rm S}_{\rm a}$ & 4.16  &  42.85  &  1.77  &  33.30  &  0.70  &  26.03  &  0.41  &  23.33  \\  
In$^{\rm S}_{\rm a}$P$^{\rm S}_{\rm a}$ & 6.46  &  58.35  &  3.76  &  35.68  &  1.36  &  25.29  &  0.82  &  24.00  \\  
In$^{\rm S}_{\rm i}$ & 3.05  &  51.51  &  1.18  &  33.05  &  0.65  &  25.96  &  0.39  &  21.89  \\  
P$^{\rm S}_{\rm i}$ & 2.95  &  40.13  &  1.39  &  24.51  &  0.69  &  20.46  &  0.45  &  16.90  \\  
In$^{\rm S}_{\rm v}$ & 11.23  &  40.15  &  6.92  &  27.99  &  4.54  &  20.10  &  2.90  &  14.65  \\  
P$^{\rm S}_{\rm v}$ & 0.92  &  26.00  &  0.49  &  18.94  &  0.36  &  16.15  &  0.24  &  14.88  \\  
In$^{\rm S}_{\rm v}$P$^{\rm S}_{\rm v}$ & 4.33  &  57.06  &  2.69  &  38.85  &  1.16  &  32.05  &  0.74  &  27.61  \\  
\end{tabular}
\end{ruledtabular}
\end{table}



\begin{figure*}
\begin{subfigure}[b]{0.49\textwidth}
\centering
\includegraphics[width=\textwidth]{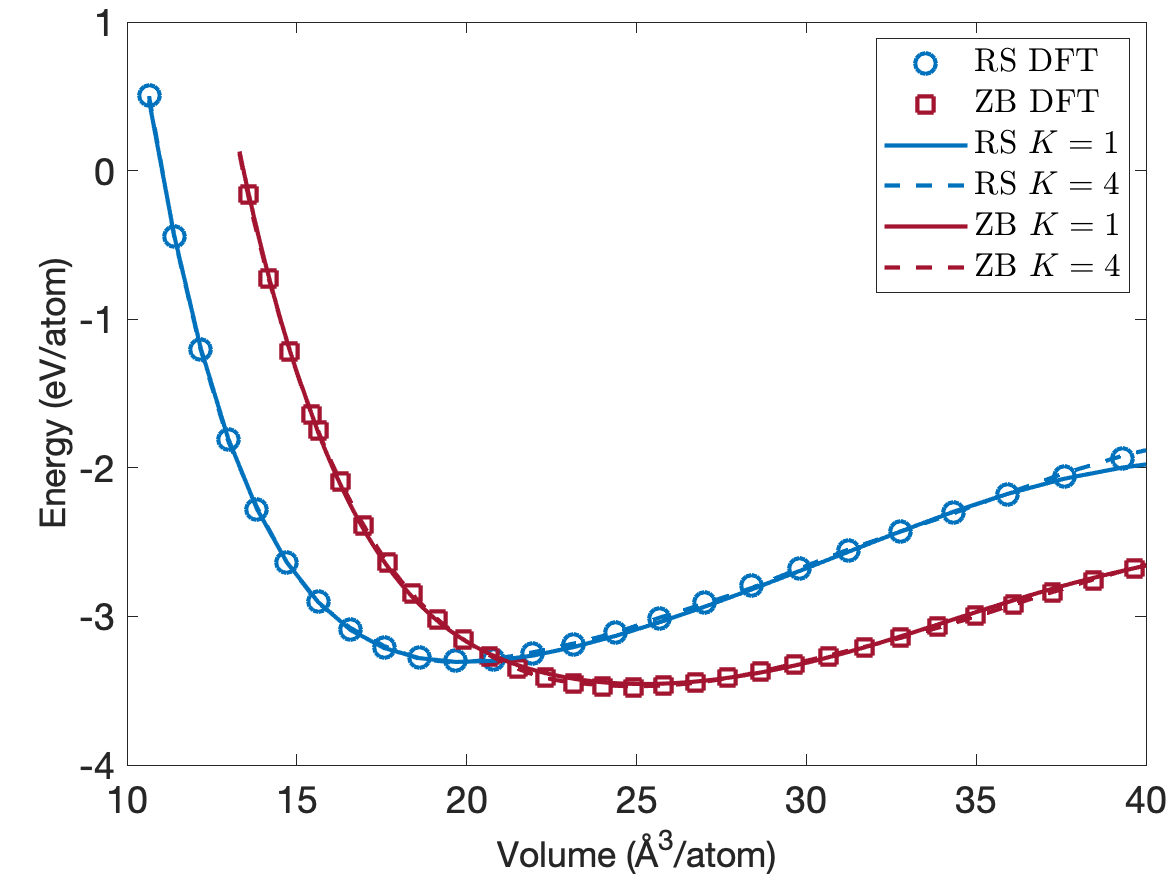}
\caption{Energy per atom versus volume per atom curve}
\end{subfigure}
\begin{subfigure}[b]{0.49\textwidth}
\centering
\includegraphics[width=\textwidth]{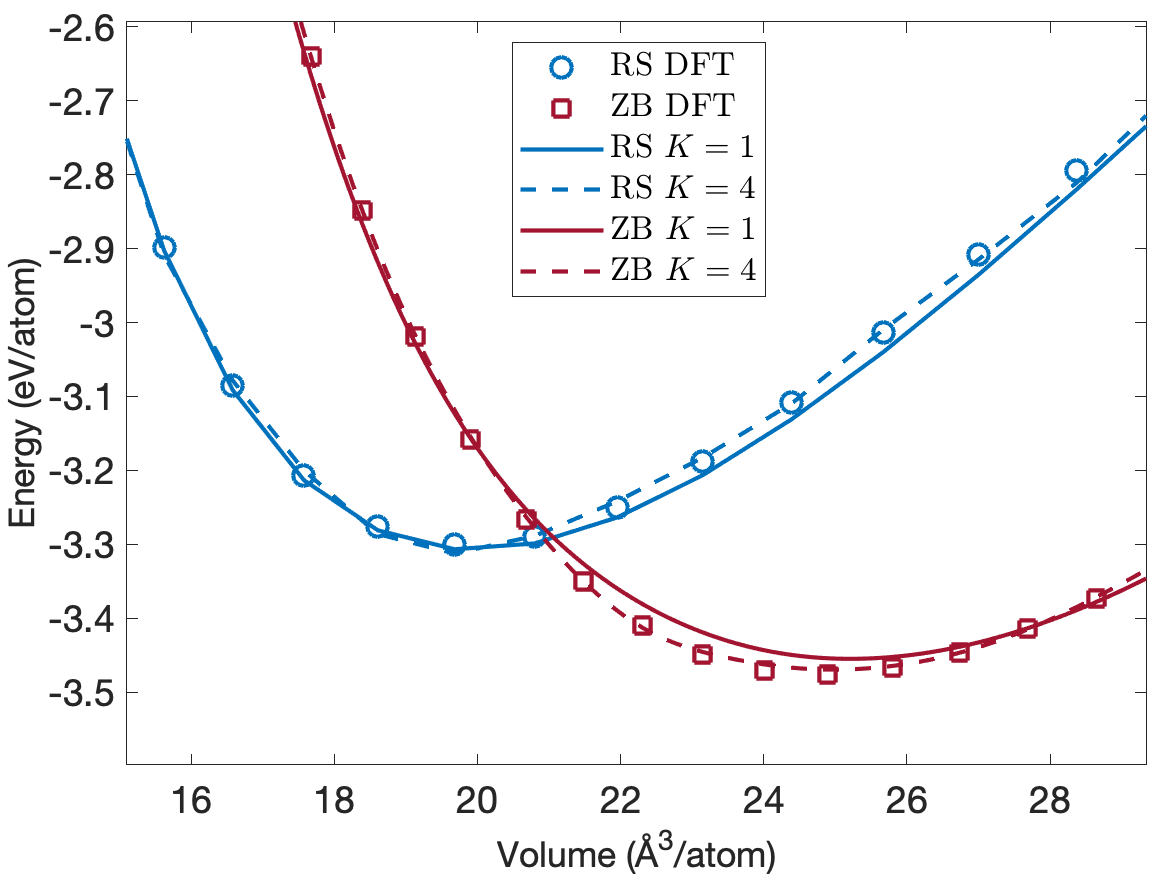}
\caption{Close-up view near the minimum energies}
\end{subfigure}
\caption{Energy per atom versus volume per atom for RS and ZB crystal structures for EAML potentials using $M=40$ descriptors in comparison with DFT data.}
\label{fig3}
\end{figure*}

\revise{One of the crucial requirements for interatomic potentials is that they predict the formation and cohesive energies  accurately.  In addition to defect formation energies, we also study cohesive energies for different low-energy crystal structures. Figure \ref{fig3} plots the energy per atom as a function of volume per atom for the rocksalt (RS) and zincblende (ZB) crystal structures. We see that the predicted cohesive energies are very close to the DFT cohesive energies for both the rocksalt (RS) and zincblende (ZB) crystal structures. Furthermore, the EAML potentials correctly predict ZB as the most stable structure and reproduce the experimental cohesive energy of -3.48 eV/atom at a volume of 24.4 $\mbox{\AA}^3$/atom \cite{Nichols1980}. The predicted cohesive energies for the RS structure match exactly the DFT value of -3.30eV/atom at a volume of 19.7 $\mbox{\AA}^3$/atom. While not plotted in Figure \ref{fig3}, the predicted cohesive energies for the wurtzite ground state structure agree well with the DFT value of -3.45eV/atom at a volume of 25.1 $\mbox{\AA}^3$/atom.}


\begin{figure*}
\begin{subfigure}[b]{0.49\textwidth}
\centering
\includegraphics[width=\textwidth]{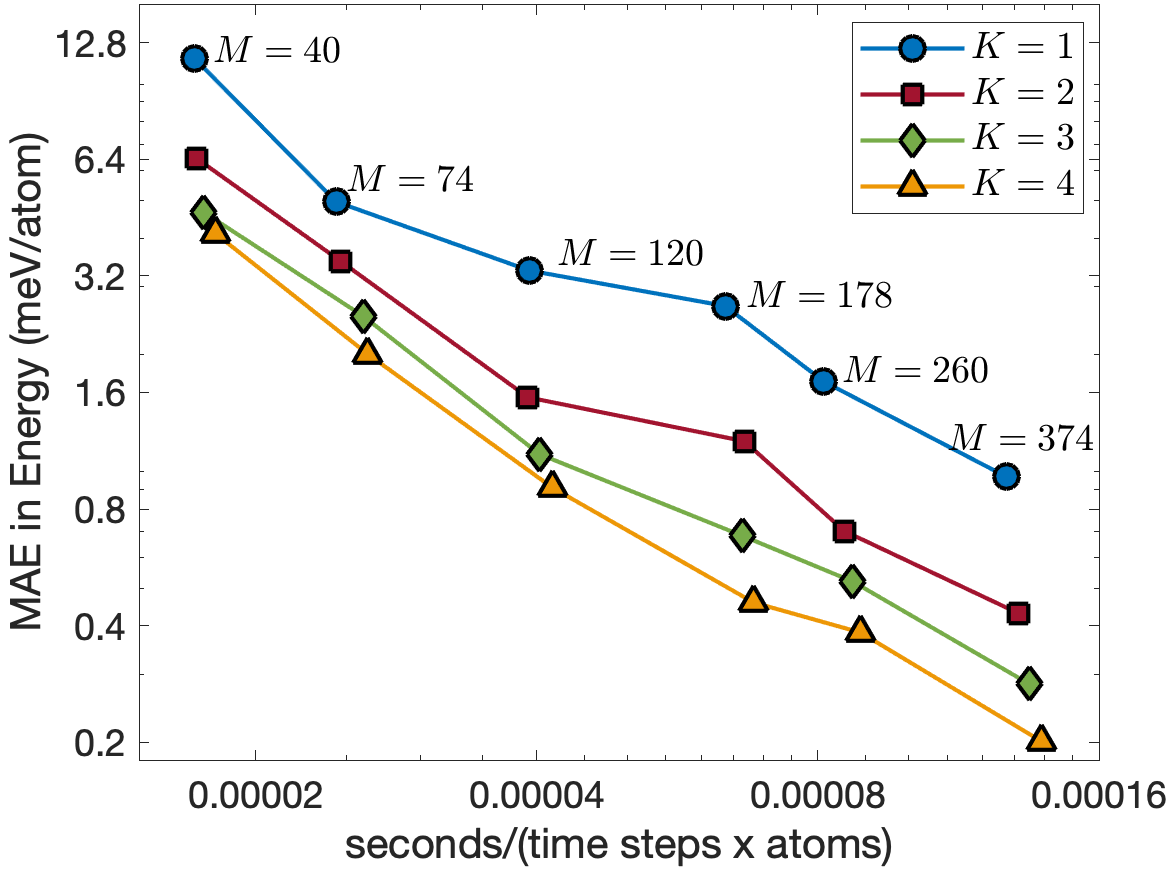}
\caption{Energy}
\end{subfigure}
\begin{subfigure}[b]{0.49\textwidth}
\centering
\includegraphics[width=\textwidth]{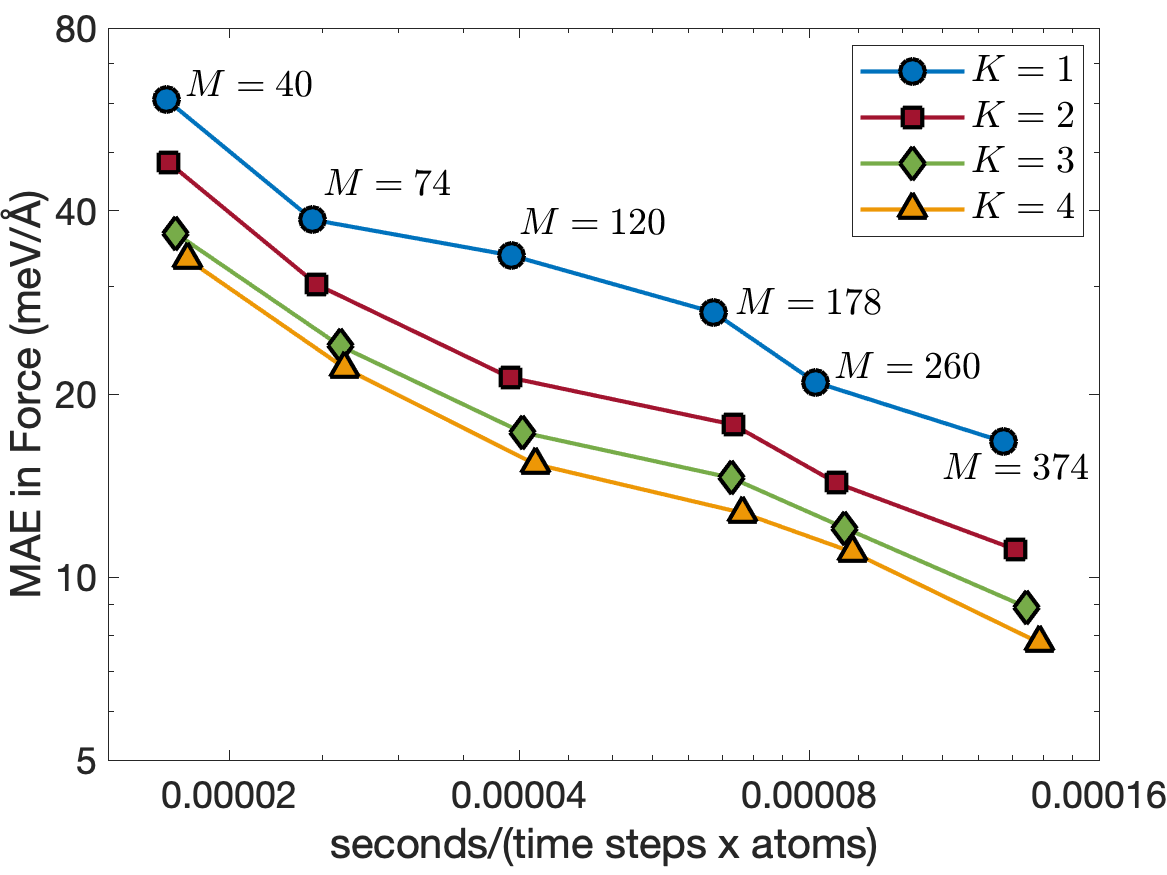}
\caption{Force}
\end{subfigure}
\caption{\label{fig4} Test errors versus the computational cost of MD simulations for the InP system of 8000 atoms. MD simulations are performed using LAMMPS \cite{Thompson2022} on a CPU core of Intel i7-1068NG7 2.3 GHz for EAML potentials with different numbers of descriptors and numbers of clusters.}
\end{figure*}

\revise{Figure \ref{fig4}
illustrates the trade-off between computational cost and accuracy for MD simulations of 8000 InP atoms performed on a single CPU core of Intel i7-1068NG7 2.3 GHz. The computational cost is measured in terms of second per time step per atom. We clearly see that the potentials for $K > 1$ are almost as fast as the potential for $K=1$ for the same number of descriptors. We also see that increasing $K$ reduces the test errors. The energy errors for $K = 4$ are about 4 times smaller than those for $K=1$, while the force errors for $K = 4$ are about 2 times smaller than those for $K = 1$. As a result, the EAML potential with $M = 120, K =4$ is more accurate and 3 times faster than the standard linear potential with $M = 374$.}

\section{Conclusions}
\label{conclusions}


\revise{We have introduced multi-element Proper Orthogonal Descriptors (PODs) for constructing machine-learned interatomic potentials. The POD descriptors incorporate elements of both internal coordinate descriptors and atom density descriptors. Our approach can be extended to arbitrary body orders and can be used to compute atom-centered symmetry functions and empirical potentials with a cost that scales only linearly with the number of neighbors. The method brings about the possibility of constructing many-body empirical potentials, while maintaining the computational cost that scales linearly with the number of neighbors. For instance, the SW and EAM potentials can be extended to include four-body terms, while atom-centered symmetry functions can be formed from the four-body POD descriptors.


We have presented an environment-decomposition method to construct accurate and transferable interatomic potentials by adapting to the local atomic environment of each atom within a system. For a dataset of $N$ atoms, atom positions and chemical species are mapped to a descriptor matrix by using the POD method. Principal component analysis (PCA) is used to reduce the dimension of the descriptor space. The $k-$means clustering scheme is applied to the reduced matrix to partition the dataset into subsets of similar environments. Each cluster represents a distinctive local environment and is used to define a corresponding local potential.  We introduce a many-body many-potential expansion to smoothly blend these local potentials to ensure global continuity of the potential energy surface. This continuity is achieved by calculating probability functions that assess the likelihood of an atom belonging to specific clusters identified within the dataset. 

We have applied the EAML potentials to Ta and InP datasets. The results show that EAML models provide significantly more accurate predictions than the standard linear model for the same number of descriptors $M$. There are several reasons behind the better accuracy of EAML potentials. First, EAML potentials have more capacity than the standard linear potential because they have a larger number of fitting coefficients (i.e., $KM$ versus $M$).  Second, owing to the probability functions that vary with the neighborhood of the central atom, EAML potentials adapt their descriptors according to the local atomic environments to capture atomic interactions more accurately than the linear potential. Third, the products of the probability functions and the descriptors contain higher body interactions than the descriptors themselves. As a result, EAML potentials can capture higher-order interactions than the linear potential. Since EAML potentials have computational complexity similar to that of the linear potential, they are more accurate and efficient. 


While PCA is used for its simplicity and straightforward implementation,  nonlinear dimensionality reduction techniques may offer some advantage. Autoencoders, Variational Autoencoders, t-Distributed Stochastic Neighbor Embedding (t-SNE) and Isomap offer the ability to uncover and preserve intricate structures in high-dimensional data that PCA might overlook. These methods are particularly useful in scenarios where the relationships among data points involve complex patterns. While $k-$means clustering is used for its simplicity and straightforward implementation,   there are several other clustering techniques that can be use to deal with very large and diverse datasets. Techniques such as hierarchical clustering, Density-Based Spatial Clustering of Applications with Noise (DBSCAN), and Gaussian Mixture Models (GMMs) offer alternative methods that can yield better clusters than $k-$means clustering.


In this paper, we consider linear regression to construct EAML models. Linear models are easy to understand and interpret because the relationship between the output and trainable parameters is linear. They are computationally inexpensive to train and require relatively low computational resources. Indeed, it takes only a few seconds to a few minutes to train EAML potentials on a personal computer. However, linear regression may be insufficient to capture complex atomic interactions without transformation of input features. Significant performance improvement can be achieved by using more sophisticated regression methods such as nonlinear regression, kernel regression, and neural networks. While these nonlinear models require much longer training times than linear models, they often yield more accurate predictions for the same computational cost \cite{Rohskopf2023}. 

}


\section*{Acknowledgements} 
We would like to thank Jaime Peraire, Robert M. Freund, Youssef Marzouk, Nicolas Hadjiconstantinou, and Spencer Wyant at MIT for the fruiful discussions on a wide ranging of topics related to this work. We would also like to thank Andrew Rohskopf, Axel Kohlmeyer and Aidan Thompson for fruitful discussions about LAMMPS implementation of POD potentials. We gratefully acknowledge the United States  Department of Energy under contract DE-NA0003965 and the the Air Force Office of Scientific Research under Grant No. FA9550-22-1-0356 for supporting this work.

\bibliography{library}

\end{document}